\documentclass[review]{elsarticle}
\usepackage{lineno}
\usepackage{subfigure}
\modulolinenumbers[5]

\usepackage[utf8]{inputenc}
\usepackage[T1]{fontenc}
\usepackage{color}
\usepackage{babel}
\usepackage{float}
\usepackage{mathtools}
\usepackage{amsmath}
\usepackage{amssymb}
\usepackage{braket}
\usepackage{graphicx}
\usepackage{hyperref}
\usepackage{microtype}
\usepackage[T1]{fontenc}
\usepackage{lmodern}
\usepackage{pdfpages}
\journal{Materials Today Physics}

\begin{document}
\tolerance=1
\emergencystretch=\maxdimen
\hyphenpenalty=10000
\hbadness=10000
\begin{frontmatter}



\title{Quantum Transport of Charge
Density Wave Electrons in Layered Materials}


\author[1]{John H. Miller, Jr \corref{cor1}}
\cortext[cor1]{Corresponding author}
 \ead{jhmiller@uh.edu}
\affiliation[1]{Department of Physics and Texas Center for Superconductivity, University of Houston, Houston, TX 77204 USA}
\author[1]{Martha Y. Suárez-Villagrán}
\author[1]{Johnathan O. Sanderson}

\begin{abstract}
The charge density wave (CDW) is a condensate that often forms in layered materials. It is known to carry electric current \emph{en masse}, but the transport mechanism remains poorly understood at the microscopic level. Its quantum nature is revealed by several lines of evidence. Experiments often show lack of CDW displacement when biased just below the threshold for nonlinear transport, indicating the CDW never reaches the critical point for classical depinning. Quantum behavior is also revealed by oscillations of period $h/2e$ in CDW conductance vs. magnetic flux, sometimes accompanied by telegraph-like switching, in $\text{TaS}_3$ rings above 77 K. Here we discuss further evidence for quantum CDW electron transport. We find that, for temperatures ranging from 9 to 474 K, CDW current-voltage plots of three trichalcogenide materials agree almost precisely with a modified Zener-tunneling curve and with time-correlated soliton tunneling model simulations.  In our model we treat the Schr\"{o}dinger equation as an emergent classical equation that describes fluidic Josephson-like coupling of paired electrons between evolving topological states. We find that an extension of this \lq classically robust' quantum picture explains both the $h/2e$ magnetoconductance oscillations and switching behavior in CDW rings. We consider potential applications for thermally robust quantum information processing systems.
\end{abstract}




\end{frontmatter}


\section{Introduction}
\label{sec:intro}
The charge density wave (CDW) is an ordered quantum fluid that often forms in layered quasi-1D or -2D materials \cite{miller2021quantum,miller2,miller3,miller1985dynamics}. Its formation can be driven by an electron (or electron-hole) condensate coupled to a softened phonon mode \cite{gruner1988dynamics,pierre,bardeen1979theory,frohlich1954theory} or by purely electronic mechanisms \cite{gerber2015ybco,gooth2019axionic,randle2021collective}. Phonons, as bosonic lattice modes, can macroscopically occupy a single quantum state that manifests itself as a lattice distortion (superlattice). This, in turn, couples to the CDW electronic condensate, often represented as an electronic charge modulation \cite{bardeen1990superconductivity,bardeen1979theory, gruner1988dynamics, pierre}.  The coherent state picture has been proposed as a framework to describe dynamical lattice modes acting on charge carriers  \cite{kim2022coherent}, and may provide a promising approach for future theoretical work. In an electron-phonon-based system, current-carrying CDW electrons transfer most of their momentum and energy to the lattice modes \cite{bardeen1979theory, frohlich1954theory}. Large condensation energies enable high transition temperatures \cite{gruner1988dynamics, pierre}, sometimes exceeding that of boiling water \cite{zybtsev2021ultra}, and large Peierls energy gaps. Temperatures for collective quantum behavior are only limited by interaction strengths and condensation energies. Superfluid and superconducting condensates in neutron stars, for example, likely have $T_c$’s of up to $\sim10^{10}$ K \cite{chamel2017superfluidity,pines1985superfluidity}. 

Some experiments \cite{tsubota2009quantum,tsubota2012aharonov} show oscillations of period $h/2e$, in CDW conductance vs. magnetic flux through $85-\mu$m circumference $\text{TaS}_3$ rings, at 5.1 K and 79 K. These provide clear examples of relatively long-range quantum coherence of CDW electrons, or perhaps electron pairs, within the condensate. In some cases, the rings also show telegraphic switching between high- and low-conductance states \cite{tsubota2012aharonov,hosokawa2015hysteretic}, suggesting transitions between distinct macroscopic states. Additional experiments, some driving the system far from equilibrium, reveal metastable \lq hidden quantum' and topological states in quasi-2D CDWs \cite{stojchevska2014ultrafast,han2015exploration,basov2017towards,zhou2021nonequilibrium,altvater2021observation}.

Several aspects of CDWs suggest their potential, both for studying the foundations of quantum physics and as platforms for quantum information processing. In the presence of dissipative normal electrons, at $\sim50$ K in $\text{NbSe}_3$ \cite{jones2000pulse}, the pulse-duration memory effect \cite{coppersmith1987pulse} is one of several natural learning phenomena \cite{gruner1988dynamics}. The CDW \lq remembers' the durations of repeated rectangular current pulses, and adjusts its voltage oscillations accordingly. Remarkably, only 1-3 pulses are needed to train the CDW, as compared to hundreds or thousands in classical simulations, sometimes no learning in classical weak-pinning simulations \cite{jones2000pulse}. This behavior suggests that CDWs might be well-suited to act as physical reservoirs in quantum reservoir computing \cite{ghosh2019quantum,tanaka2019recent}. Here a material, acting as a reservoir with quantum properties, is placed between input and output neural network layers to enhance learning speeds. 

CDW-based devices may also hold potential for quantum computing at temperatures much higher than 15 mK, typical for superconducting quantum computers \cite{arute2019quantum,blais2021circuit}. A CDW with a complete Peierls gap acts as a nonlinear, nearly harmonic oscillator when cooled \cite{mihaly1988low}. Its ac response, up to several tens of GHz, would enable higher frequency operation of a microwave qubit device similar to the transmon. Its large Peierls energy gap would tend to reduce quasiparticle poisoning of quantum coherence. Macroscopic occupation of collective modes might further enhance the operating temperature, possibly well above milli-Kelvin temperatures in our proposed CDW-superconductor hybrid devices \cite{John_patent_2021,John_patent_2022}, to be discussed here.  Finally, the CDW may be an ideal platform to help elucidate the boundary between the quantum and classical worlds.

\begin{figure}[H]
\centering
\includegraphics[scale=0.32]{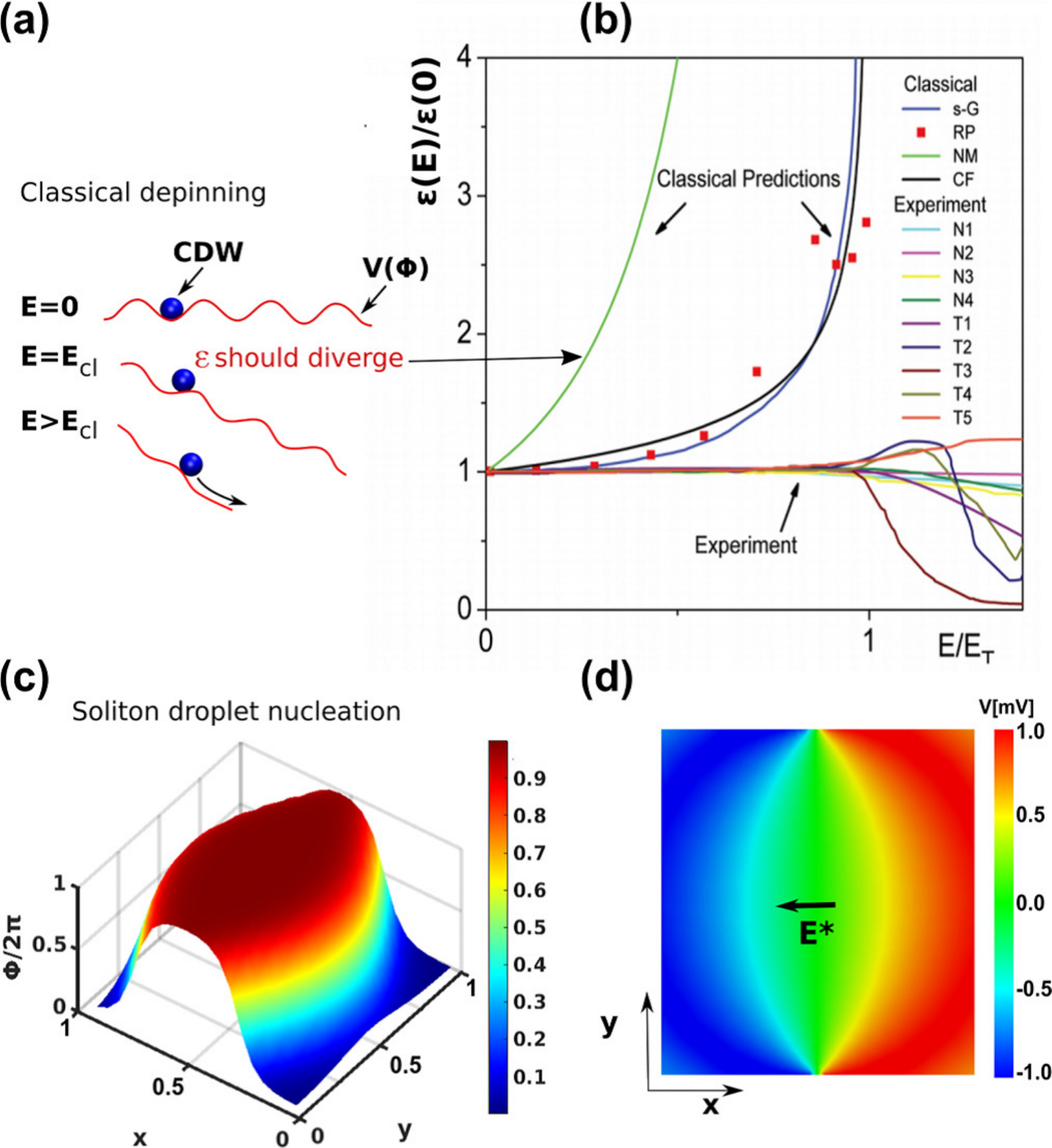}
\caption{\label{figure1} \label{electrostatic_potential} (a) Classical model of CDW depinning in a tilted washboard pinning potential. (b) Bias-dependent dielectric response, showing classical predictions vs experiment. Classical models include classical sine-Gordon (s-G); random pinning (red squares, RP: Ref.\protect\cite{littlewood1986sliding}); renormalization group (NM; Ref.\protect\cite{narayan1994avalanches}); $|f|^{-2}$; and incommensurate harmonic chain (CF; Ref.\protect\cite{coppersmith1988threshold}), $|f|^{-0.34}$, models, where $f=1-E/E_T$. Some $\mathrm{NbS}\mathrm{e}_{3}$ measurements were carried out in our laboratory \protect\cite{miller3} using a bridge circuit ($\mathrm{NbS}\mathrm{e}_{3}$: N1, 45 K, 1 kHz; N2, 120 K, 3 kHz; N3, 45 K, 100 kHz), while additional measurements were carried out by ZG (Ref.\protect\cite{zettl1984phase}; N4 $\mathrm{NbS}\mathrm{e}_{3}$, 42 K, 3.2 MHz) and WMG (Ref.\protect\cite{wu1985dielectric}; $\mathrm{TaS_{3}}$: T1, 130 K, 5 MHz; T2, 100 K, 1 kHz; T3, 110 K, 1 kHz; T4, 100 K, 1 kHz; T5, 100 K, 10 kHz). (c) Fluidic soliton domain wall pairs formed by aggregates of nucleated soliton dislocations. They surround a bubble of lower energy within which the CDW phase has advanced by $2\pi$. Here $x$ represents the CDW chain direction. (d) The gradient in electrostatic potential (red = positive, right side; blue = negative, left side), due to charged soliton droplet pairs, creates an internal electric field $E^{\ast}$.\protect\cite{miller3} Adapted from Ref. \protect\cite{miller2021quantum}.}
\end{figure}

A quasi-1D CDW modulates the charge along each chain, $\rho_{i}(x,t)=\rho^{0}_{i}(x,t)+\rho_{1}\text{cos}[2k_{F}x-\phi_{i}(x,t)]$, where $k_{F}$ is the Fermi wavevector \cite{pierre, gruner1988dynamics}. A phase deformation over several wavelengths carries net charge per length per chain $(2e/2\pi)(\partial\phi/\partial x)$, which couples to and produces electric fields. In the pinned CDW, a $\pm 2 \pi$ phase deformation on a single chain acts as a soliton of charge $\pm 2e$, with boson-like properties. In our model, many such deformations among parallel chains can coalesce into fluidic domain walls (Fig.\ref{figure1}) or soliton liquids\cite{matsuura2015charge}, or form arrays \cite{altvater2021observation}. We interpret these as dynamical condensates away from equilibrium, emergent quantum states, which grow and diminish with time. The threshold electric field $E_{T}$ for soliton pair creation is a Coulomb blockade effect due to the internal electric field generated by the soliton-antisoliton pair \cite{miller2021quantum, miller2,miller3}. This is usually much smaller than the classical depinning field $E_{cl}$ for \lq sliding.' The quantum picture is supported by NMR \cite{ross1986nmr} and X-ray diffraction \cite{requardt_1998}  experiments showing little CDW phase displacement just below the threshold for nonlinear transport. This is corroborated by flat dielectric response vs. bias field \cite{miller3}, as illustrated in Fig. \ref{figure1}, and flat harmonic mixing response vs. bias below threshold \cite{miller1985dynamics}.  A phase diagram \cite{miller2,miller3} shows the conditions under which quantum pair creation or classical depinning are expected to occur. In cases where quantum transport dominates, the CDW remains near the bottom of the pinning potential well and never reaches the critical point for classical depinning.

According to our model, quantum transport of CDW electrons occurs via fluidic, Josephson-like coupling between evolving states that correspond to different charging energies \cite{miller2021quantum,miller2,miller3}. This is similar to time-correlated single electron tunneling \cite{averin1986coulomb}, but involves a macroscopic number of electrons. The process, however, should not be viewed as tunneling of a single macroscopic object, but rather as coherent flow of many microscopic entities within the quantum fluid. The Schrödinger equation describing coupling between macrostates is treated as an emergent classical equation \cite{feynman} – robust up to the transition temperature, $\sim474$ K in one case \cite{zybtsev2021ultra}. 

In this paper we compare reported CDW $I-V$ and conductance curves, for $\text{NbS}_3$, $\text{TaS}_3$, and $\text{NbSe}_3$ over the temperature range 9 - 474 K, with a modified Zener-tunneling curve proposed by Bardeen \cite{bardeen1980tunneling}. We carry out similar comparisons with time-correlated soliton tunneling (ST) simulations \cite{miller2021quantum,miller2,miller3}. We extend the model to interpret $h/2e$ magneto-conductance oscillations and switching behavior in $\text{TaS}_3$ rings \cite{tsubota2009quantum,tsubota2012aharonov}. Finally, we discuss possible applications in quantum information processing \cite{John_patent_2021,John_patent_2022}. 

\section{Results and discussion}

In the time-correlated soliton tunneling (ST) model \cite{miller2021quantum,miller2,miller3}, narrow band noise and voltage oscillations result from time-correlated, Josephson-like tunneling of microscopic entities between successive charging energy macrostates \cite{miller2021quantum}. The simulations yield unparalleled agreement between theory and experiment \cite{miller2,miller3,miller2021quantum}. Current voltage $(I-V)$ characteristics are computed by averaging over several cycles, and also exhibit excellent agreement \cite{miller2,miller3,miller2021quantum}  with experiments on $\mathrm{NbSe_{3}}$.  For a wide range of parameters, the computed $I-V$ curves are found to match Bardeen's modified Zener-tunneling curve \cite{bardeen1980tunneling}:  

\begin{equation}
I_{CDW}=G_{max}[V-V_{Tm}]\mathrm{exp}\left[-\frac{V_{0}}{V}\right].\label{eq:1}
\end{equation}

\noindent The apparent \lq measured' threshold $V_{Tm}$ is often found, in simulations, to be larger than the Coulomb blockade threshold $V_{T}$ of the ST model \cite{miller2,miller3,miller2021quantum}, particularly in cases where $V_{0}/V_{T}>>1$ and the $I-V$ curves are rounded. On the other hand, it is found that the two are about equal when $V_{0}/V_{T}<1$, corresponding to nearly piecewise linear $I-V$ plots. 

\subsection{Coherent tunneling behavior in CDW \textit{I-V} characteristics}

\begin{figure}[H]
\centering
\includegraphics[scale=0.3]{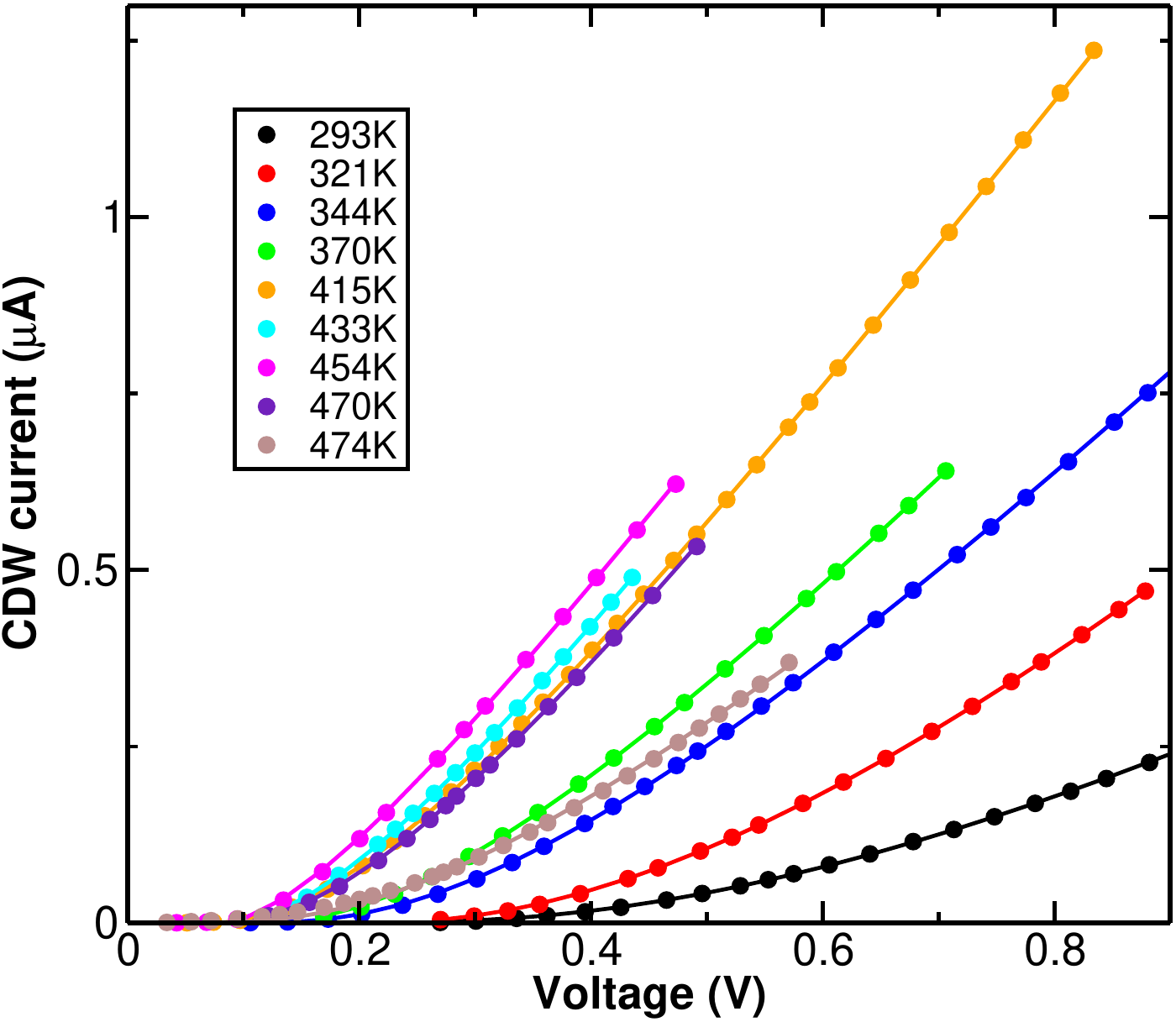}
\caption{\label{figure2} Experimental (solid circles) vs. quantum theoretical Bardeen-Zener (solid lines) $I-V$ curves for $\mathrm{NbS_{3}}$ at various temperatures ranging from room temperature up to 474 K. The experimental plots were extracted from reported differential conductance data for sample $\#4$  \cite{zybtsev2021ultra}, as discussed in the text.}
\end{figure}

We first use the Bardeen-Zener (BZ) formula (Eq. \ref{eq:1}) to analyze reported \cite{zybtsev2021ultra} CDW transport data on  $\mathrm{NbS_{3}}$ for temperatures up to 474 K. We scanned the differential conductance data for sample $\#4$, without RF irradiation, in Fig. 2 of ref. \cite{zybtsev2021ultra}, and subtracted the normal conductance. We then integrated to obtain $I-V$ curves, which were compared to the BZ curve of Eq. \ref{eq:1}. The results are shown in Fig. \ref{figure2}, and use the parameters shown in Table \ref{table1}. The $I-V$ plots in Fig. \ref{figure2} show nearly precise agreement between the BZ curve and the experimental data. The results thus strongly support a quantum mechanism of CDW transport. 

\begin{table}[H]
\caption{\label{table1} BZ formula (Eq. \ref{eq:1}) parameters used to fit to fit the experimental data for $\mathrm{NbS_{3}}$ \cite{zybtsev2021ultra} in Fig. \ref{figure2} at various temperatures.}
\begin{center}
\begin{tabular}{|c|c|c|c|}
\hline 
Temperature (K) & $G_{max} (\mathrm{M}\Omega)^{-1}$ & $V_{Tm}$ (V)& $V_0$ (V)\tabularnewline
\hline 
\hline 
293 & 1.005 & 0.222& 0.942\tabularnewline
\hline 
321 & 1.636 & 0.234& 0.708\tabularnewline
\hline 
344 & 1.680 &0.132& 0.452\tabularnewline
\hline 
370 & 1.780 &0.133& 0.329\tabularnewline
\hline 
415 & 2.181 &0.082& 0.237\tabularnewline
\hline 
433 & 2.214 &0.101& 0.181\tabularnewline
\hline 
454 & 2.166 &0.092& 0.132\tabularnewline
\hline 
470 & 2.130 &0.108& 0.209\tabularnewline
\hline 
474 & 1.453 &$1*10^{-5}$& 0.469\tabularnewline
\hline
\end{tabular}
\end{center}
\end{table}

One can collapse the data onto a single line using the following relation:

\begin{equation}
\mathrm{ln}(g) \equiv \mathrm{ln}\left[ \frac{I_{CDW}}{G_{max}[V-V_{Tm}]}\right] = -\frac{V_o}{V}.\label{eq:2}
\end{equation}

\noindent This is carried out for the $\mathrm{NbS_{3}}$ sample $\#4$ data discussed above, and for sample $\#6$ \cite{zybtsev2021ultra}, in Fig. \ref{figure3}(a), for values $V/V_{Tm} > 2$. The plot clearly shows that the points collapse to a single straight line, for temperatures ranging from 293 K (room temperature) up to 474 K. We have performed a similar analysis of reported $\mathrm{TaS_{3}}$ \cite{zettl26} and $\mathrm{NbSe_{3}}$ \cite{bardeen1990superconductivity} data. The results are shown in Figs. \ref{figure3}(b) and (c), while the cumulative sets of data are compiled in Fig. \ref{figure3}(d). The data for all three materials converges to a single straight line, consistent with the BZ curve and therefore indicating quantum transport.

\subsection{Soliton tunneling (ST) model simulations}
We have also carried out simulations using the time-correlated ST model \cite{miller2,miller3,miller2021quantum}, with focus on $\mathrm{NbS_{3}}$ \cite{zybtsev2021ultra}. In the ST model, the threshold field $E_{T}$ is a Coulomb blockade effect due to the internal field generated by nucleated soliton pairs, similar to a parallel plate capacitor.  In the current-carrying state, a dimensionless \lq vacuum angle'  \cite{coleman1976more} $\theta$ is introduced [inset, Fig.\ref{figure4}] in terms of the evolving displacement charge $Q$ using $\theta=\frac{2\pi Q}{Q_{0}}$. Here $Q_{0}=2Ne$ is the charge of a fluidic soliton domain wall, where $N$ is the number of parallel chains. The time-correlated ST model \cite{miller2,miller3,miller2021quantum} includes a shunt resistance $R$, representing normal, uncondensed electrons, in parallel with a capacitive tunnel junction depicting soliton tunneling. When the phase has advanced to $<\phi>=2\pi n$ where $C=\frac{\epsilon A}{l}$, as in single electron tunneling \cite{averin1986coulomb}, the voltage is proportional to net displacement charge: 

\begin{eqnarray}
V & = & \frac{1}{C}\left(Q-nQ_{0}\right)=\frac{Q_{0}}{2\pi C}\left(\theta-2\pi n\right).\label{eq:3}
\end{eqnarray}

\begin{figure}[H]
\centering
\includegraphics[scale=0.25]{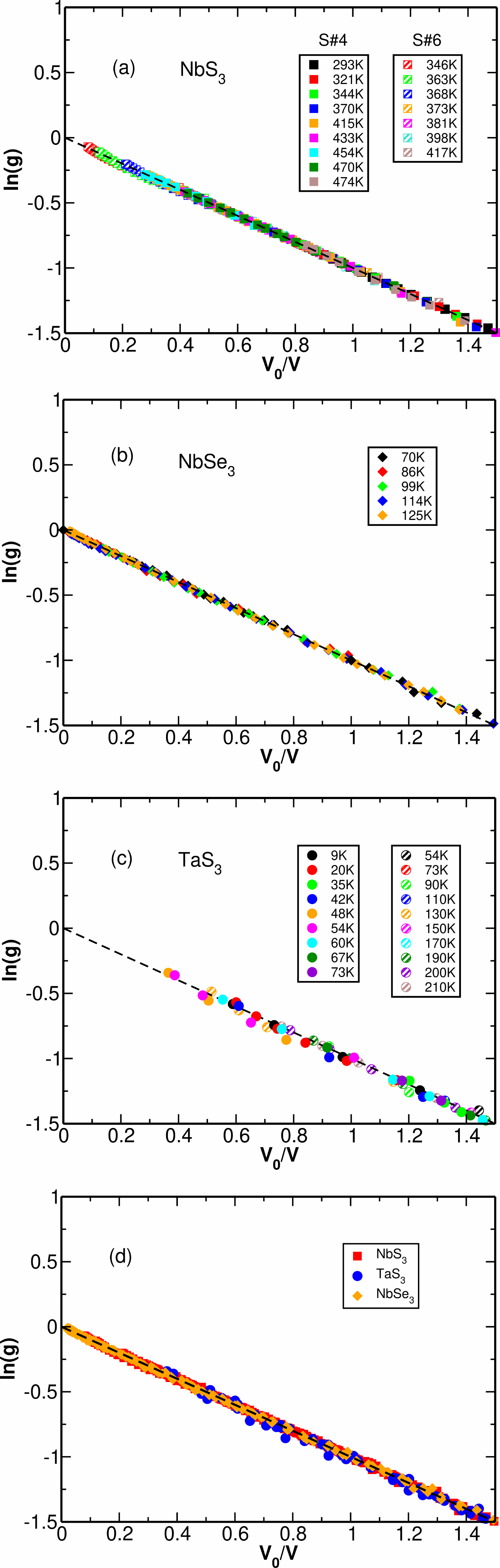}
\caption{\label{figure3} Plots of ln(g) vs. $V_{0}/V$, Eq. \ref{eq:2}, for: (a) $\mathrm{NbS_{3}}$ \cite{zybtsev2021ultra}, (b) $\mathrm{NbSe_{3}}$ \cite{bardeen1990superconductivity}, and (c) $\mathrm{TaS_{3}}$ \cite{zettl26}. Panel (d) shows a cumulative plot for all three materials. See Supplementary Information for parameters used.}
\end{figure}

\noindent  

\noindent More generally the voltage is
$V=\left(\frac{Q_{0}}{2\pi C}\right)\left[\theta-<\phi>\right]$.
This results in a total current $I=I_{n}+I_{CDW}$, where: $I_{n}=\left(\frac{Q_{0}}{2\pi RC}\right)\left[\theta-<\phi>\right]$
is the normal current and: $I_{CDW}=\frac{dQ}{dt}=\left(\frac{Q_{0}}{2\pi}\right)\left[\frac{d\theta}{dt}\right]$
is the CDW current.

Using $I_{CDW}=I-I_{n},$ and defining $\omega=\frac{2\pi I}{Q_{0}}$
and $\tau=RC$, yields the following equation for the $\theta$ time
derivative, proportional to CDW displacement current \cite{miller2,miller3,miller2021quantum}:
\begin{eqnarray}
\frac{d\theta}{dt} & = & \omega-\frac{1}{\tau}\left[\theta-<\phi>\right].\label{eq:4}
\end{eqnarray}

\noindent We compute $<\phi>$ by solving the Schr\"{o}dinger equation: 

\begin{eqnarray}
i\hbar\frac{d\Psi_{0,1}}{dt} & = & U_{0}\Psi_{0,1}+T\Psi_{1,0}.\label{eq:5}
\end{eqnarray}

\noindent This describes Josephson-like coupling, via the tunneling matrix element $T$, from one macrostate to the next. Here $\Psi_{0}(t)$ and $\Psi_{1}(t)$ depict the original and emerging macrostate amplitudes for the system to be in the pinning potential wells centered at $<\phi>\sim0$ (black solid parabola in the inset to Fig. \ref{figure3}), and $<\phi> \sim2\pi$ (green dashed parabola), respectively.  These are treated as classically robust order parameters. As the system evolves, the macrosates $\Psi_{n}$ and $\Psi_{n+1}$,  centered near $\phi \sim 2\pi n$ and $2\pi (n+1)$, respectively [Fig. \ref{figure4} inset], become coupled by the tunneling matrix element as each parabola crosses the next. Following Feynman \cite{feynman}, Eq. \ref{eq:5} is viewed as an emergent \textquotedblleft classical\textquotedblright{} equation, since many microscopic processes occur coherently. 

\begin{figure}[H]
\centering
\includegraphics[scale=0.55]{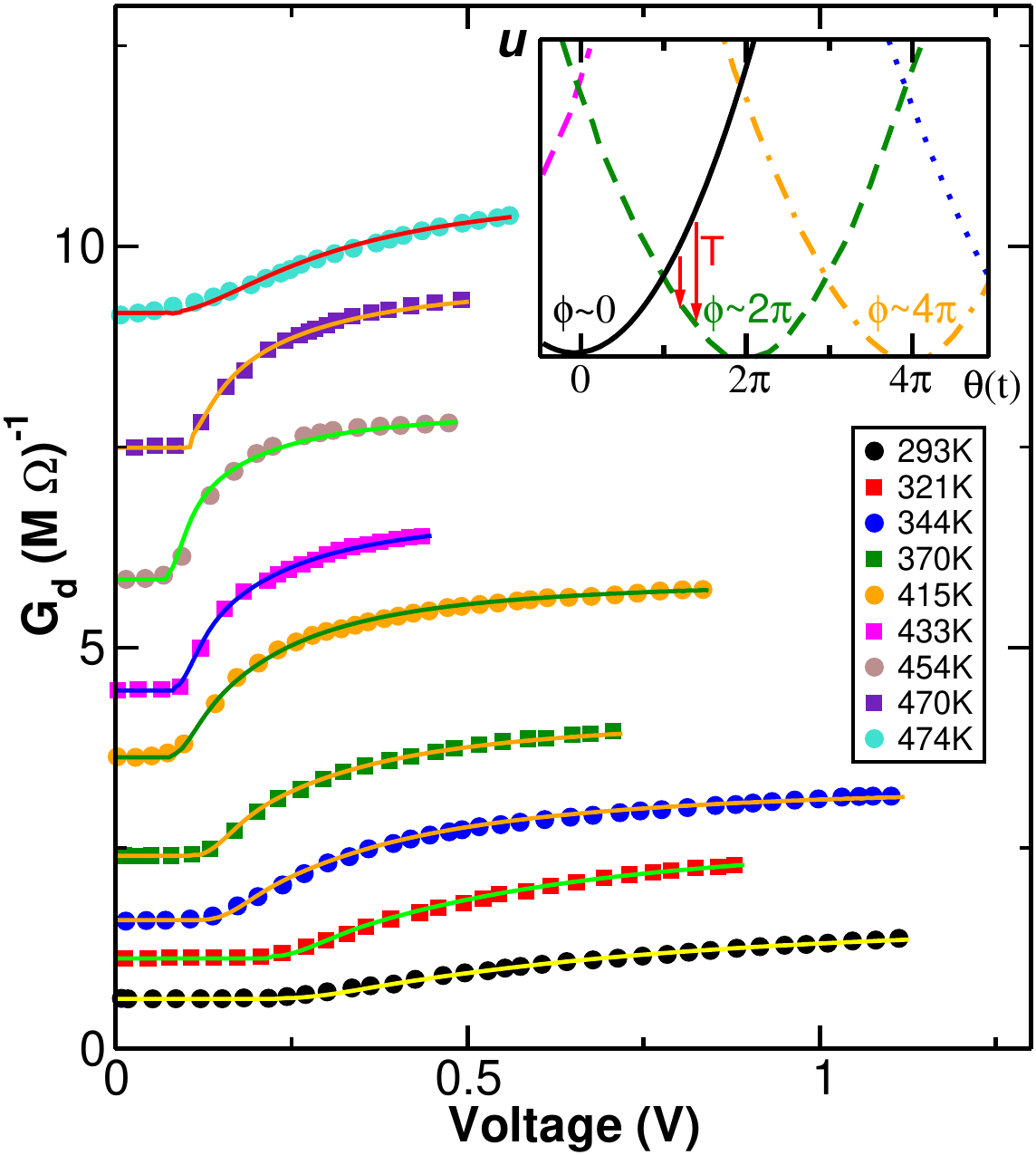}
\caption{\label{figure4} Experimental differential conductance data, $G_{d}$, for $\mathrm{NbS_{3}}$ sample $\#4$  \cite{zybtsev2021ultra} (solid symbols), as compared to time-correlated ST model simulations (text and inset).}
\end{figure}

As one charging energy branch crosses the next, for $\theta > \pi , \space 2\pi$, etc., the driving force $F$ becomes proportional to the energy difference between adjacent parabolas in the inset to Fig. \ref{figure4} \cite{miller2,miller3,miller2021quantum} $F=4\pi u_{E}\theta'_{n}$,
where $\theta'_{n}=\theta-2\pi\left(n+\frac{1}{2}\right)$. An advance in phase is accompanied by soliton pair production at the boundaries, so the matrix element $T$ is estimated \cite{bardeen1961tunnelling} to have a Zener-like form \cite{miller2,miller3,miller2021quantum}, equivalent to Schwinger pair production \cite{cohen2008schwinger} $T(F)=-4F\lambda\exp\left[\frac{-F_{0}}{F}\right]$\textcolor{black}{,
where $\lambda$ is a length scale and $F_{0}$ is proportional to the Zener field $E_{0}$. The negative energy in the bubble between soliton pairs is balanced by their positive energy, so $T$ couples
states of equal energy: $U_{0}=U_{1}=U$. Defining}\textcolor{red}{{}
}$\Psi_{0,1}(t)=\chi_{0,1}\exp\left[-\frac{iUt}{\hbar}\right]$ then simplifies the Schr\"{o}dinger equation to: $\frac{i\hbar\partial\chi_{0,1}}{\partial t}=T\chi_{1,0}$.

The Schr\"{o}dinger equation, together with Eq. \ref{eq:4}, are converted
to dimensionless form employing parameters defined in Table S4 of the Supplementary Information and in Ref. \cite{miller2021quantum}. This
yields the following coupled differential equations: 

\begin{equation}
\frac{dc_{1}}{dt'}=\left[\gamma q'_{n}\exp\left(-\frac{q_{0}}{q'_{n}}\right)\right]c_{0},\:\mathrm{for}\:q'_{n}>0,\label{eq:6}
\end{equation}

\noindent and

\begin{equation}
\frac{dc_{0}}{dt'}=-\left[\gamma q'_{n}\exp\left(-\frac{q_{0}}{q'_{n}}\right)\right]c_{1},\:\:\mathrm{for}\:q'_{n}>0.\label{eq:6-1}
\end{equation}

\noindent These are integrated numerically with initial values $c_{0}=1$
and $c_{1}=0$, yielding $<\phi>=2\pi[n+p]$, where $p=\left|c_{1}\right|^{2}$.
The transition from branch $n$ to $n+1$ is considered complete when $p$ exceeds a cutoff close to $1$. For each value of total current, the CDW voltage and current are averaged over several cycles to yield dimensionless $I-V$ plots. These are scaled to the experimental data using measured normal conductance values $G_{n}$ and scaling voltages, $V_{norm}$, which are usually equal to the measured threshold voltages, and finally converted to differential conductance.

The resulting theoretical plots, using parameters in Table \ref{table2}, are shown in Fig. \ref{figure4} in comparison with the experimental $\mathrm{NbS_{3}}$ data from sample $\# 4$ \cite{zybtsev2021ultra}. As with the BZ formula comparisons in Figs. \ref{figure2} and \ref{figure3}, the ST model simulations show excellent agreement with the experimental $\mathrm{NbS_{3}}$ data \cite{zybtsev2021ultra}, for temperatures ranging from 293 K (room temperature) up to 474 K.

\begin{table}[H]
\caption{\label{table2} Parameters used in the ST model simulations.}
\begin{center}
\begin{tabular}{|c|c|c|c|c|}
\hline 
Temperature (K) & $\gamma$ & $q_0$ &Vnorm (V)& gnorm $(\mathrm{M}\Omega)^{-1}$ \tabularnewline
\hline 
\hline 
293 & 2.475 & 5.40&0.222&0.624\tabularnewline
\hline 
321 & 2.13 & 3.9&0.234&1.127\tabularnewline
\hline 
344 & 1.55 &4.5&0.132&1.603\tabularnewline
\hline 
370 & 1.09 &3.5&0.133&2.406\tabularnewline
\hline 
415 & 0.88 &4.1&0.082&3.632\tabularnewline
\hline 
433 & 0.75 &2.9&0.101&4.465\tabularnewline
\hline 
454 & 0.54 &2.4&0.092&5.851\tabularnewline
\hline 
470 & 0.42 &2.80&0.108&7.490\tabularnewline
\hline 
474 & 0.235 &18&0.03&9.170\tabularnewline
\hline
\end{tabular}
\end{center}
\end{table}

\subsection{Inclusion of quantum mechanical phase in ST model}
In our previous work \cite{miller2021quantum,miller2,miller3,miller2021quantum} and above ST model simulations, we fixed the relative phase (phase difference) between macrostates to $\pi/2$, and defined $\chi_0=c_0$ and $\chi_1=ic_1$. Here we relax that assumption to include the effects of quantum mechanical phase variations. Now defining $\chi_0=r_0e^{i\delta_0}$ and $\chi_1=r_1e^{i\delta_1}$, and letting $T'=-T/\hbar$, since $T(F)$ is negative \cite{miller2021quantum,miller2,miller3}, leads to the following equations, after some algebraic manipulation and defining $\delta = \delta_1 - \delta_0$:
\begin{equation}\label{eq:r0}
    \frac{1}{r_0}\frac{dr_1}{dt}=T'\text{sin}\delta,
\end{equation}

\begin{equation}\label{eq:delta1}
    \frac{d\delta_1}{dt}=T'\frac{r_0}{r_1}\text{cos}\delta,
\end{equation}

\begin{equation}\label{eq:r1}
    \frac{1}{r_1}\frac{dr_0}{dt}=-T'\text{sin}\delta,
\end{equation}
and 

\begin{equation}\label{eq:delta0}
    \frac{d\delta_0}{dt}=T'\frac{r_1}{r_0}\text{cos}\delta.
\end{equation}
Subtracting Eq. (\ref{eq:delta0}) from Eq. (\ref{eq:delta1}) yields:
\begin{equation}\label{eq:delta}
    \frac{d\delta}{dt}=T'\left[\frac{r_0}{r_1}-\frac{r_1}{r_0}\right]\text{cos}\delta
\end{equation}

\noindent The above phases represent quantum mechanical phase differences between macrostates. Note the similarity to the Josephson current-phase relations in Eqs.(\ref{eq:r0}) and (\ref{eq:r1}). The time derivative $dr_1/dt$ correlates with CDW current and Eq.(\ref{eq:r1}) reveals the growth in amplitude for the next charging energy branch, $n+1$. Eq.(\ref{eq:r0}) shows the decrease in amplitude for branch $n$. Equation (\ref{eq:delta}) describes convergence of the system toward the optimum phase that maximizes the current. This will be $\delta=\pi/2$ in the absence of other factors such as an applied magnetic flux.

\subsection{Modeling of CDW ring behavior}
To interpret the $h/2e$ Aharonov-Bohm (AB) oscillations in CDW rings\cite{tsubota2009quantum,tsubota2012aharonov}, we extend the ST model to allow for spatial variations in quantum mechanical phase and vector potential.  These efforts suggest that a course-grained model based on stacked Josephson-junction (JJ) superconducting quantum interference devices (SQUIDs) \cite{lewandowski1991dc,konopka1996equilibrium,krasnov2002stacked}, explains not only the AB oscillations but, at least qualitatively, the telegraph switching behavior. We incorporate multiple CDW domains on the two branches of the SQUID-like ring to simulate CDW current vs. magnetic flux in CDW rings \cite{tsubota2009quantum,tsubota2012aharonov}. Recall that, in the presence of an applied magnetic field, the vector potential around the ring satisfies $\oint \vec{A}\cdot d\vec{l}=\Phi$. We then use the extended time-correlated soliton tunneling model to simulate the time-evolution, where each segment, or domain, has its own evolving matrix element $T_{j,k}$ and phase, $\delta_{j,k}$ \cite{lewandowski1991dc,konopka1996equilibrium,krasnov2002stacked}.

In the previous sections we modeled the entire CDW as a single “tunnel junction” with a single composite matrix element ($T$ or $T'$). Here we extend the model to allow for spatial variations in magnetic vector potential. This will enable interpretation of the $h/2e$ AB oscillations in CDW rings\cite{tsubota2009quantum,tsubota2012aharonov}. Our preliminary efforts indicate that a course-grained model based on stacked Josephson-junction (JJ) superconducting quantum interference devices (SQUIDs) \cite{lewandowski1991dc,konopka1996equilibrium,krasnov2002stacked}, explains not only the AB oscillations but, at least qualitatively, the telegraph switching behavior. In order to simulate CDW current vs. magnetic flux for the CDW ring experiments \cite{tsubota2009quantum,tsubota2012aharonov}, we incorporate multiple CDW domains on the two branches of the SQUID-like ring. Recall that, in the presence of an applied magnetic field, the vector potential around the ring satisfies $\oint \vec{A}\cdot d\vec{l}=\Phi$. We then use the extended time-correlated soliton tunneling model to simulate the time-evolution, where each segment, or domain, has its own evolving matrix element $T_{j,k}$ and phase, $\delta_{j,k}$ \cite{lewandowski1991dc,konopka1996equilibrium,krasnov2002stacked}.

\begin{figure}[H]
\centering
\includegraphics[scale=0.5]{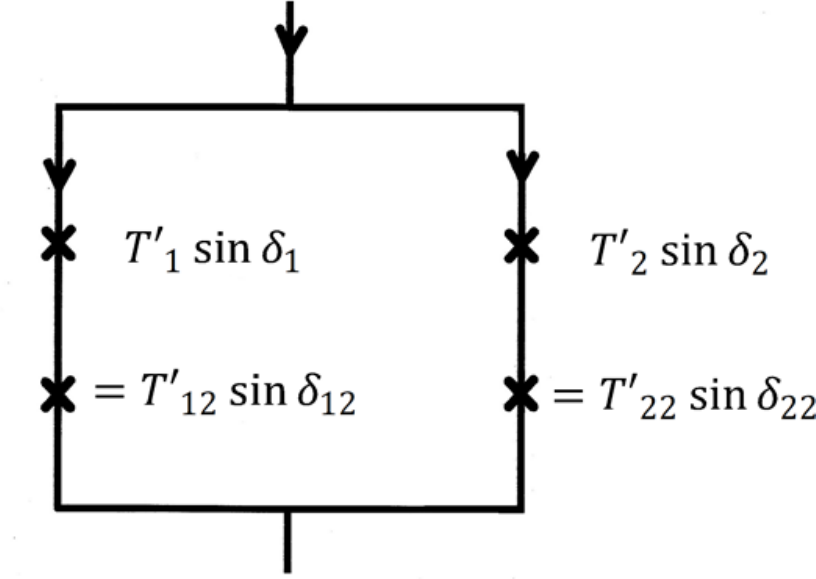}
\caption{\label{figure5}Simplified four-junction model of a CDW ring.}
\end{figure}

\begin{figure}[H]
\centering
\includegraphics[scale=0.4]{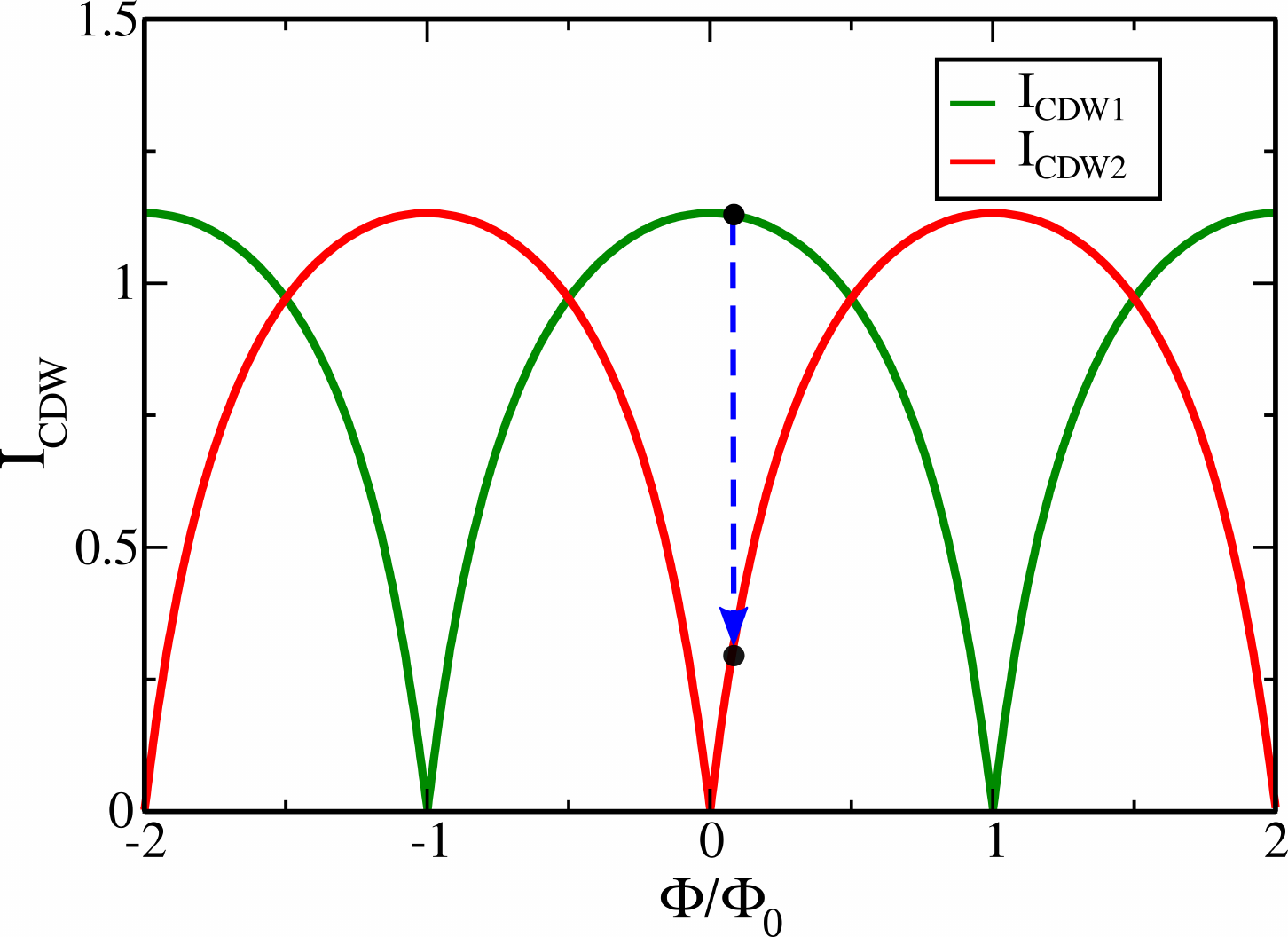}
\caption{\label{figure6} Results using a four-segment version of the time-correlated soliton tunneling model, showing multivalued CDW current vs. normalized magnetic flux, $\Phi/\Phi_0$, where $\Phi_0 = h/2e$. The blue dashed arrow illustrates a telegraph-like transition from the green, $\text{ICDW}_1$, to the orange-red, $\text{ICDW}_2$,  branch.}
\end{figure}

A four-segment circuit is sufficient to qualitatively explain both the $h/2e$  oscillations \cite{tsubota2009quantum} and telegraph-like switching noise suggesting transitions between different states \cite{tsubota2012aharonov}. We draw upon work \cite{lewandowski1991dc, konopka1996equilibrium,krasnov2002stacked}, on series-parallel JJ arrays, also called stacked JJ SQUIDs. Here We follow Lewandowski’s \cite{lewandowski1991dc} and Konopka et al.’s \cite{konopka1996equilibrium} treatment of a four-junction SQUID, with two JJs in series on each of two parallel ring branches (Fig. \ref{figure5}). Since $I_{CDW}$ is the same for all junctions in series on a given ring branch (provided $I_{CDW} \gg I_{normal}$), the phases adjust themselves such that phase of the weakest junction, with the lowest critical current, becomes the driving phase. Thus, the phases in Eqs. \ref{eq:r0} and \ref{eq:r1} adjust themselves such that $r^{(1,2)}_0$, $r^{(1,2)}_1$, and their time derivatives are conserved on a given ring branch (1 or 2) in Fig. \ref{figure5}, despite temporal variations.  We assume that, at any given time, $T'_{1}\leq T'_{12}$ and $T'_{2}\leq T'_{22}$. Eq. (\ref{eq:r0}) is modified to the forms:

\begin{equation}\label{eq:newr0}
    \frac{1}{r^{(1)}_0}\frac{dr^{(1)}_1}{dt}=T'_{1}\text{sin}\delta_1=T'_{12}\text{sin}\delta_{12}
\end{equation}
and 

\begin{equation}\label{eq:newr1}
    \frac{1}{r^{(2)}_0}\frac{dr^{(2)}_1}{dt}=T'_{2}\text{sin}\delta_2=T'_{22}\text{sin}\delta_{22}.
\end{equation}
If we assume that the relevant charge for Josephson-like tunneling is $2e$ then, as in the stacked JJ SQUID \cite{lewandowski1991dc,konopka1996equilibrium}, the phases will adjust themselves to satisfy: $\delta_1 + \delta_{12} - \delta_2 - \delta_{22} = 2\pi \Phi/\Phi_0$, where $\Phi_0 = h/2e$. Any differences in $T'$ will result from variations in the prefactor, Zener field, and threshold field. Figure \ref{figure6} shows an example of results, assuming equal $T'$ values, on computed, time averaged CDW current (above threshold) vs. magnetic flux. Note the multivalued nature, as in a stacked JJ SQUID, of the CDW current. The blue dotted arrow illustrates a telegraph-like transition between the two (green and orange-red) branches.

\subsection{Potential applications in quantum information processing}

Perhaps the greatest impact of CDWs might be their ability to overcome the \lq tyranny of temperature' in quantum computing. Existing superconducting quantum computers \cite{arute2019quantum,blais2021circuit} must be cooled to about 15 mK. This is problematic, not only due to the large dilution refrigerator, but also to the need for three coaxial lines per qubit (including one for flux tuning) and associated microwave components to transmit signals to and from room temperature (RT) electronics. These include (RT to mK) attenuators and (mK to RT) circulators and a chain of low-noise preamplifiers. A variety of CDW-superconductor hybrid devices have been proposed by one of the authors to enable operation at higher temperatures \cite{John_patent_2021,John_patent_2022}. One of the simplest is a nonlinear $LC$ circuit (Fig. \ref{figure7}), in which the roles of linear and nonlinear elements are reversed vs. the transmon \cite{arute2019quantum,blais2021circuit}. Due due its enormous nonlinear dielectric response \cite{gruner1988dynamics}, a CDW crystal at low (but not necessarily mK) temperatures can play the role of nonlinear capacitor, as shown on the right side of Fig. \ref{figure7}.

When cooled, the nonlinearity of the $LC$ circuit enables unequal spacing between quantized energy levels, as discussed in the next section. The ground and excited states, for example, can be designated as $\ket{0}$ and $\ket{1}$ states. Due to the bosonic nature of phonons, we hypothesize that these states can become macroscopically occupied. Approximating the coupled macrostates as a two-level system, their time evolution can be described by treating the Schr\"{o}dinger equation as an emergent classical equation, similar to our previous discussion. Readout and coupling to the nonlinear resonator can be similar to techniques used for the transmon. Dispersive coupling and readout, where the qubit couples to a linear resonator detuned from the qubit, has emerged as a method to avoid strongly perturbing the qubit  \cite{blais2021circuit}. In dispersive readout, the state of the qubit, $\ket{0}$ or $\ket{1}$, shifts the resonance frequency of the coupling resonator. This affects its reflection coefficient when probed with microwave electronics. A key difference here is that, in the proposed CDW hybrid device, the input and output signals can be substantially larger than those used for the transmon.

Key signatures of quantum behavior, for the device in Fig. 7, would be Rabi oscillations between the $\ket{0}$ and $\ket{1}$ states \cite{blais2021circuit}. To probe such behavior, one could capacitively or inductively coupled to the device and apply a weak electromagnetic signal of amplitude proportional to $\Omega$ at a fixed frequency $\omega$. In the next section, we treat the amplitudes, $c_0$ and $c_1$, of the $\ket{0}$ and $\ket{1}$ macrostate occupancies, as classically robust order parameters, as in our previous discussions. Following the approach used for transmons, we employ a two-level approximation and consider a periodic signal of frequency $\omega$ and amplitude proportional to $\Omega$, which induces transitions between the two states separated by energy $\hbar \omega_0$ per mode. Subtracting off a global energy shift \cite{blais2021circuit}, the Hamiltonian maps onto that for a spin-1/2 particle in the presence of dc and ac magnetic fields:

\begin{equation}
\hat{H}=-\frac{\hbar\omega_0}{2}\hat{\sigma}_z-\frac{\hbar \Omega}{2}\text{cos}(\omega t)\hat{\sigma}_x .
\end{equation}

\noindent As discussed in the next section, the time-dependent  Schr\"{o}dinger equation yields two coupled differential equations for the macrostate amplitudes $c_0$ and $c_1$. These are solved using the rotating wave approximation \cite{shore1993jaynes,irish2022defining,blais2021circuit}. On resonance, when $\omega = \omega_0$, this yields the simplest result for Rabi oscillations: $\left| c_0 \right|^2=\text{cos}^2\left(\Omega t / 4 \right)$ and $\left| c_1 \right|^2=\text{sin}^2\left(\Omega t / 4 \right)$. The off-resonance case, $\omega \neq \omega_0$, is discussed in the next section. The collective nature of such Rabi oscillations might be similar to coherent many-body Rabi oscillations \cite{dudin2012observation, dourado2021coherent} or super-Rabi oscillations \cite{bin2021parity}. An open questions, if observed, is whether they would be smooth, as suggested by our classical treatment of Schr\"{o}dinger time evolution. Perhaps more exciting would be probabilistic telegraph-like switching behavior, where the Rabi oscillations are reconstructed by doing repeated measurements at various time intervals. This possibility is suggested by the switching behavior observed in CDW rings \cite{tsubota2012aharonov}. The latter case might be indicative of processes similar to Dicke superradiance and superabsorption in collections of two-level systems coupled to a common electromagnetic field \cite{dicke1954, higgins2014}.

\subsection{Potential for Rabi oscillations in CDW-superconductor hybrid devices}
Rabi oscillations would be key signatures of quantum behavior in the hybrid device depicted in Fig. \ref{figure7}. Here we model the time evolution of coupled macrostates by using the 2-level approximation. Subtracting off a global shift in energy \cite{blais2021circuit}, the Hamiltonian for a driven 2-level system describing each mode can be modeled as a spin-$1/2$ particle affected by dc and ac magnetic fields:

\begin{equation}
\hat{H}=-\frac{\hbar\omega_0}{2}\hat{\sigma}_z-\frac{\hbar \Omega}{2}\text{cos}(\omega t)\hat{\sigma}_x .
\end{equation}

\noindent Here $\omega_0$ is the qubit resonance frequency and $\Omega$ is proportional to the amplitude of the driving signal. We wish to determine how the state, 
\begin{equation}
    \ket{\Psi(t)}=\begin{pmatrix} c_0(t) \\ c_1(t)  \end{pmatrix},
\end{equation}

\noindent of the 2-level system evolves subject to the initial condition 

\begin{equation}
    \ket{\Psi (0)} =\begin{pmatrix} 1 \\ 0  \end{pmatrix}.
\end{equation}

\noindent The system evolves via the time-dependent Schrödinger equation:
\begin{equation}
    i\hbar\frac{d}{dt}\ket{\Psi(t)}=\hat{H}\ket{\Psi(t)} .
\end{equation}

\noindent This yields two coupled differential equations:

\begin{equation}
    i\frac{dc_0}{dt} = - \frac{\omega_0}{2}c_0 - \frac{\Omega}{2} \text{cos}(\omega t) c_1 ,
\end{equation}

\noindent and

\begin{equation}
    i\frac{dc_1}{dt} = \frac{\omega_0}{2}c_1 - \frac{\Omega}{2} \text{cos}(\omega t) c_0 .
\end{equation}

\noindent As in our previous discussions, for the collective system we treat the amplitudes $c_0(t)$ and $c_1(t)$ as classically robust macrostates in the above equations. Eliminating diagonal terms using the interaction picture and employing the rotating wave approximation \cite{shore1993jaynes,irish2022defining,blais2021circuit}, yields the following results for the time evolution of the macrostates:

\begin{equation}
    c_0(t)=e^{i\omega t/2}\left[ \text{cos}\left(\frac{1}{2}\omega't\right)-i\frac{\Delta \omega}{\omega'}\text{sin}\left(\frac{1}{2}\omega't\right)\right]
\end{equation}

\begin{equation}
    c_1(t)=ie^{-i\omega t/2}\left[ \frac{\Omega}{2\omega'}\text{sin}\left(\frac{1}{2}\omega't\right)\right]
\end{equation}

\noindent where $\omega'=\sqrt{\Delta \omega^2+\Omega^2/4}$ and the detuning between the driving and resonance frequency is $\Delta \omega =\omega-\omega_0$. On resonance, when the detuning is zero, this yields the maximum Rabi oscillations between the two macrostates:

\begin{equation}
    \left| c_0 \right|^2=\text{cos}^2\left(\frac{\Omega t}{4} \right)
\end{equation}

\noindent and

\begin{equation}
    \left| c_1 \right|^2=\text{sin}^2\left(\frac{\Omega t}{4} \right) .
\end{equation}

\subsection{Simplified analysis of energy levels for CDW-superconductor qubit}

\begin{figure}[H]
\centering
\includegraphics[scale=0.9]{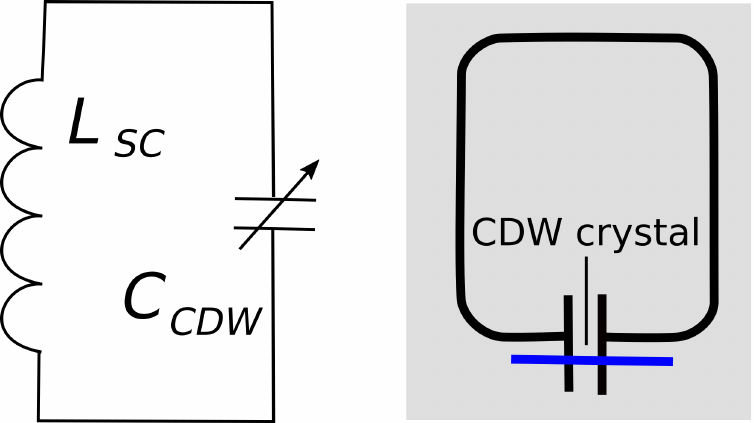}
\caption{\label{figure7} Left. Schematic of a nonlinear \textit{LC} resonator in which the roles of linear and nonlinear elements are reversed vs. the transmon qubit. The right side shows an example of a resonator in which the nonlinear dielectric properties of a CDW crystal (blue) forms a nonlinear capacitor, while the inductor consists of a superconducting loop patterned on a substrate.
The huge dielectric response of the CDW, with $\epsilon_r$ ranging from $10^5$ to $10^8$ along the chain direction, enables it to dominate
the capacitance across the gap.}
\end{figure}

The analysis of energy levels for a transmon is discussed in a review article on circuit quantum electrodynamics \cite{blais2021circuit}.  Flux and charge play the roles of “position” and “momentum” respectively. These roles, as are the roles of voltage and current, are reversed here, where charge $q$ acts as “position,” flux $\varphi$ as “momentum,” and inductance $L$ as “mass.” In operator form, the linear part of the Hamiltonian for a single mode of the CDW-superconductor qubit (e.g., Fig. \ref{figure7}) can then be written as \cite{John_IEEE_2023}:

\begin{equation}\label{hamiltonian_single_mode_CDW}
\hat{H}=\frac{\hat{\varphi}^2}{2L}+\frac{1}{2}L\omega^2_0\hat{q}^2,
\end{equation}

\noindent where $\omega_0=\frac{1}{LC}$ and where $L$ and $C$ are the inductance and limiting low-amplitude capacitance relevant to a single mode.
For a CDW in a linear chain compound with $N$ condensed electrons, its kinetic inductance due to inertial response of the coupled electron-phonon system is: $L_k^{(total)}=\frac{M^{*}l^2}{Ne^2}$. Here $l$ is the distance between contacts and the effective mass $M^*=M_F+m_e$ incorporates the electron mass $m_e$ and the large Fröhlich mass $M_F$ due to transfer of kinetic energy from CDW electrons to lattice vibrations (phonons) \cite{Bardeen_PRB_1989}. The distance $l$ is related to the number $N_1$ of condensed CDW electrons in a single chain by $l=N_1 \lambda_c$, where $\lambda_c$ is comparable to the CDW wavelength.

Noting that the number of parallel chains is given by, $N_{ch}=\frac{N}{N_1}$, the CDW kinetic inductance can be rewritten as: $L_k^{(total)}=\frac{M^*\lambda_cl}{N_{ch}e^2}$. The kinetic inductance of a single chain is then given by: $L_k^{(1)}=\frac{M^*\lambda_cl}{e^2}$. Counterintuitively, the total inductance scales inversely with $N_{ch}$. This is related to the fact that inductive reactance, like resistors in parallel, goes inversely with number of inductors in parallel. The effective inductance per mode, when coupled to an external superconducting inductor, is expected to scale similarly. By contrast, the total capacitance, like capacitors in parallel, will scale with $N_{ch}$. The frequency $\omega_0$ will thus be the same whether one refers to the single- or multi-mode values of inductance and capacitance.

Considering a single mode in this simplified analysis, by analogy to the quantum harmonic oscillator, the charge and flux operators are related to the creation and annihilation operators by: 
\begin{equation}
    \hat{q}=\sqrt{\frac{\hbar}{2L\omega_0}}\left( \hat{a}+\hat{a}^{\dag} \right)
\end{equation}
and 
\begin{equation}
    \hat{\varphi}=i\sqrt{\frac{\hbar L\omega_0}{2}}\left( \hat{a}^{\dag}-\hat{a} \right) .
\end{equation}
These obey the commutation relation, $\left[\hat{q},\hat{\varphi}\right]=i\hbar$, whereas $\left[\hat{a},\hat{a}^{\dag} \right]=1$. The linear portion of the Hamiltonian, Eq. (\ref{hamiltonian_single_mode_CDW}), then becomes:

\begin{equation}\label{H_reduce}
    \hat{H}^{(0)}=\hbar \omega_0 \left[\hat{a}^{\dag} \hat{a}+\frac{1}{2}\right].
\end{equation}
Equation (\ref{H_reduce}) leads to equally spaced energy levels: $E_n^{0}=\hbar \omega_0\left[n+\frac{1}{2}\right]$.

A cooled CDW behaves as a nonlinear dielectric, whose dielectric response changes with the amplitude of the applied electric field, and hence voltage, as well as with frequency. This leads to generation of harmonics \cite{sherwin1984chaotic} and/or enables harmonic mixing signals \cite{miller1985dynamics} when one or more oscillatory signals are applied. Since displacement charge scales with voltage, assuming decreasing dielectric response with electric field \cite{miller1985dynamics}, the discussion below assumes a negative quartic nonlinear term.  However, the device will function with either positive or negative nonlinearities, simply by removing the minus sign in the shifts in energy levels for the positive case. Here we treat the quartic nonlinearity as a perturbation, which can be expressed in terms of a dimensionless parameter, $\alpha<1$. Higher order terms can be included but are not necessary to demonstrate unequally spaced energy levels. The nonlinearity is expressed in terms of displacement charge $q$, which scales with voltage.

Defining $q_0 \equiv \sqrt{\frac{\hbar}{ L\omega_0}}$, the nonlinear perturbation to the Hamiltonian can be written (for the negative case) as: 
\begin{equation}
    \hat{H}^{(1)}=\alpha \delta \hat{H}=-\alpha \hbar \omega_0\left(\frac{\hat{q}}{q_0}\right)^4=-\alpha \frac{\hbar \omega_0}{4}\left(\hat{a}+\hat{a}^{\dag}\right)^4 .
\end{equation}

\noindent Carefully taking the commutation relations into account, the lowest-order correction to the ground state energy becomes: $E_0^{(1)}=-\alpha\left(\frac{\hbar \omega_0}{4}\right)\bra{0}\left(\hat{a}+\hat{a}^{\dag}\right)^4\ket{0}=-\alpha\left(\frac{3}{4}\right)\hbar\omega_0$. The ground state energy is thus: $E_0(\alpha)=\hbar\omega_0 \left(\frac{1}{2}-\left(\frac{3}{4}\right)\alpha+O\left(\alpha^2\right) \right)$. Following a similar procedure for higher excited states, 
$E_n^{(1)}=-\alpha \left(\frac{\hbar\omega_0}{4}\right)\bra{n}\left(\hat{a}+\hat{a}^{\dag}\right)^4\ket{n}$, we find that $E_n^{(1)}=-\alpha \left(\frac{\hbar\omega_0}{4}\right)\left[3+6n+2n^2\right]$. This yields the result:
\begin{equation}
    E_n(\alpha)=\hbar \omega_0 \left[ n+\frac{1}{2}- \frac{\alpha}{4}\left(3+6n+2n^2\right)+O\left(\alpha^2 \right) \right].
\end{equation}
The spacing between energy levels then becomes: 
\begin{equation}\label{space_energy}
    E_{n+1}-E_n=\hbar\omega_0\left[1\mp (n+2)\alpha\right],
\end{equation}
where the $+$ sign would apply for the case of a positive, rather than negative, quartic nonlinear term.  In CDWs, the quartic nonlinearity can result from an expansion of the periodic pinning energy, analogous to the Josephson coupling energy. In ferroelectrics and quantum paraelectrics, other mechanisms are involved. The specific origin of the nonlinearity, however, is not critical to the device’s functionality so such materials could also play the role of nonlinear capacitor.

The $n$-dependence of the energy difference in Eq. (\ref{space_energy}) clearly shows that the additional nonlinearity provided by a nonlinear dielectric material leads to unequally spaced energy levels. This enables one to designate two states to represent a qubit, three to represent a qutrit, or more generally $d$ states to designate a qudit.   The quality factor $Q$ of the nonlinear $LC$ resonator able to differentiate the qubit, or qutrit, etc., states from others would need to approximately satisfy: $Q>\frac{1}{\alpha}$. Certain CDW materials, such as $(\text{TaSe}_4)_2\text{I}$, show harmonic response indicating significant nonlinearity \cite{sherwin1984chaotic}, suggesting the $Q$ values may not need to be as high as required for transmons in some cases. This opens the possibility of using a cooled normal metal inductor, instead of a superconductor, for the nonlinear $LC$ resonator in cases where $\alpha$ is large enough.

\section{Conclusions and Future Directions}\label{discussion}

The above results build on a significant body of evidence \cite{bardeen1990superconductivity,miller2,miller3,miller2021quantum,tsubota2009quantum, tsubota2012aharonov} indicating that CDW electron transport is an extraordinary collective quantum phenomenon. There is overwhelming evidence proving that electrons, as well as every other known particle in the universe, behave according to the laws of quantum physics \cite{kleppner2000}. This fact alone means that \lq sliding' can hardly be considered a fundamental theory of electron transport.

The evidence specific to CDWs, moreover, shows that the sliding picture is inadequate even as a macroscopic model. This evidence includes: 1) lack of sufficient CDW displacement to reach the critical point for classical depinning \cite{ross1986nmr,requardt_1998,miller2021quantum,miller2,miller3}; 2) $I-V$ curves showing Zener tunneling-like behavior \cite{bardeen1979theory,bardeen1980tunneling,bardeen1990superconductivity,thorne_1985}; 3) more subtly, complete mode locking with an ac source at frequencies orders of magnitude above the dielectric relaxation frequency \cite{thorne_1987}; 4) linear (admittance) and nonlinear (mixing) ac responses that agree with a modified version of photon assisted tunneling theory \cite{miller1985dynamics,miller2,miller3,miller_these}; 5) extremely rapid learning phenomena \cite{jones2000pulse}; 6) narrow band noise spectra and coherent voltage oscillation shapes that quantitatively agree with time-correlated ST model simulations \cite{miller2021quantum,miller2, miller3};  and 7) most compellingly, Aharonov-Bohm oscillations of period $\hbar/2e$, both in   $\text{NbSe}_3$ crystals with columnar defects \cite{Latyshev1997} and in $\text{TaS}_3$ rings \cite{tsubota2009quantum,tsubota2012aharonov}. The latter behavior is only seen for the CDW, and \emph{not} the normal electrons, indicating that electrons within the CDW condensate show greater quantum coherence.

The results in this paper further support the quantum behavior of CDW electrons. They reveal the highest known temperatures for any Earth-bound system for cooperative quantum charge transport, enabled by large condensation and Peierls gap energies. Higher temperature collective quantum phenomena, such as superconductivity and superfluidity in neutron stars \cite{chamel2017superfluidity, pines1985superfluidity, sedrakian2019superfluidity}, are enabled by even larger condensation energies but require extreme gravitational pressures not accessible in Earth-bound applications. 

The time-correlated ST model \cite{miller2,miller3,miller2021quantum}, in our view, represents a significant improvement over Bardeen's phenomenological picture \cite{bardeen1979theory,bardeen1980tunneling,bardeen1985macroscopic}. As a macroscopic theory, the ST model treats the CDW as a quantum fluid that flows coherently between charging energy macrostates. These macrostates are represented as complex order parameters, as in Feynman's Josephson effect derivation \cite{feynman} and in the London \cite{London} and Ginzburg-Landau \cite{Ginzburg,bardeen1962critical} theories. Another advantage of the ST model is that it provides a clear mechanism, essentially the backflow of normal electrons, for dissipation. In Fig. \ref{figure4}, for example, both the normal and CDW conductances are encompassed by the ST model simulations.  A microscopic theory, however, is still needed to advance this important branch of condensed matter physics.  Such a theory would potentially benefit physics as a whole, by enhancing our understanding of the emergence of classical reality from the underlying quantum substrate.

The ST model in its present form is agnostic about several key issues. These include: (1) the detailed nature of the evolving topological states, e.g., fluidic soliton domain walls, quasi-1D domain wall arrays \cite{bardeen1985macroscopic}, 2D dislocation arrays \cite{altvater2021observation}, etc.; (2) the microscopic charge involved in tunneling, whether $e$, $2e$, or fractional charge; and (3) the detailed nature of the coherent tunneling process, presumably with a Zener-like matrix element, e.g., Josephson-like or even actual Josephson tunneling of paired electrons. A uniform CDW consists of a condensate of electron-hole pairs coupled to a condensate of $2k_F$ phonons \cite{bardeen1979theory}, and is electrically neutral. An applied electric field thus couples to charged kinks, fluidic domain walls, or dislocations that emerge from a uniform ground state. On a per electron basis, the effective gap extracted from the Zener field $E_0$, e.g., for a typical TaS$_3$ sample, is minute, a few tens of $\mu \text{eV}$ \cite{bardeen1980tunneling}, orders of magnitude smaller than $kT$. This underscores the need to develop a theory based on coherent transport of many particles within the condensate. A key idea is that, like the BCS energy gap in a superconductor, the Peierls energy gap, large compared to $kT$, prevents thermal excitations while allowing fluidic transport within the condensate \cite{bardeen1979theory, bardeen1980tunneling, bardeen1985macroscopic, bardeen1990superconductivity}. 

One possibility, at least for certain materials and temperature ranges, is actual Josephson tunneling of paired carriers between fluidic domain walls and anti-domain walls, e.g., in Fig. 1. There may be a pairing of electrons or holes within each domain wall and, in fact, the charge of each single-chain soliton is $\pm 2e$ in a fully condensed system. Although a charged, fluidic domain wall would have superfluidic properties, superconductivity in the overall system would be prevented by dissipation due to interaction with the normal carriers. Intriguingly, when normal carriers are frozen out at low temperatures, certain materials show nearly dissipationless transport in the $I-V$ plots \cite{mihaly1988low}. A future microscopic description of the superfluidic droplets might employ an approach similar to that for pairing in the doped Mott insulator \cite{phillips2020exact} or for the treatment of pair density waves \cite{chen2023widespread}.

A more complete theory could, alternatively, adopt a  non-perturbative quantum field theoretic approach. This could build a many-chain quantum sine-Gordon (sG) model that includes interactions between parallel chains and between charged kinks. The (1+1)-D sG model has been studied extensively  \cite{coleman1975quantum,floreanini1987self,Rajaraman,nitta2015non}, and shown to transform between bosonic sG and fermionic, massive Thirring representations. Recently, it has been studied for its relevance to quantum computing \cite{roy2019quantum,wang2019quantum}. When electrostatic interactions are included, the sG model becomes a variant of the massive Schwinger model \cite{coleman1976more}, which includes the vacuum angle $\theta$. Future studies, for example, could use nonequilibrium dynamical mean-field theory \cite{aoki2014nonequilibrium,suarez2020two} to study a quantum dynamical phase transition \cite{heyl2018dynamical} in the region $\theta \gtrsim \pi$ just above the CDW threshold field. This is also relevant to deconfinement and CP breaking at $\theta=\pi$ in Yang-Mills theories and QCD \cite{chen2020deconfinement}.

The massless Thirring Hamiltonian, representing a single spin-chain of an unpinned CDW, can be diagonalized \cite{maiti1991theory}. The Fermi velocity ends up being replaced by the phason velocity $c_{0}=(m/M^*)^{1/2}v_{F}$, while fermions, known as quantum solitons, emerge with charge $e^*=(m/M^*)^{1/4}e$.  Here $M^*=m+M_F$, where $M_F$ is the Fr\"{o}hlich mass \cite{frohlich1954theory}.  In the limit $M^{*}/m\rightarrow 1$, they become equivalent to electrons near the right and left Fermi wavevectors, $\pm k_F$. The topological charge per spin-chain, however, is $e$ for a pinned, fully condensed CDW.  The resulting massive Thirring model then includes strong fermion-fermion interactions \cite{maiti1991theory}, due to electron-phonon coupling in the Fröhlich model \cite{frohlich1954theory}. Counting both spins in a fully condensed  CDW, the topological charge of a single-chain soliton dislocation becomes $2e$. The above arguments highlight the importance of determining the relevant charge of each microscopic entity involved in quantum transport.

A microscopic understanding of CDW dynamics could ultimately enable thermally robust quantum information processing that does not require cooling with a dilution refrigerator. Example concepts include the proposed CDW - superconductor hybrid qubit devices discussed here. Also of relevance are the switching effects \cite{tsubota2012aharonov} seen in $h/2e$ magnetoconductance oscillations in CDW rings \cite{tsubota2009quantum}, which may also be interpreted in terms of fractional charge \cite{hosokawa2015single}. This suggests non-Abelian character \cite{nitta2015non} of CDW solitons, which could enable another form of topological quantum computing \cite{sola2020majorana,nayak2008non,stern2013topological}.

Quantum transport of topological entities, in particular flux vortices, is also relevant to applications employing high-$T_c$ superconductors (HTSs). Their high upper critical fields, for example, have been proposed to be exploited in designs of more compact fusion reactors employing high-field HTS magnets \cite{creely2020overview}. A magnetic Weber blockade effect, similar to the Coulomb blockade effect discussed here, has been proposed in a model of time-correlated vortex tunneling in superconductors \cite{miller2017time}. The model accounts for a plateauing effect observed in thickness dependent critical currents ($I_c$'s) of HTS tapes \cite{selvamanickam2009high}, and why the thickness dependent $I_c$ improves with the use of ${\mathrm{CeO_2}}$ spacer layers \cite{foltyn2007materials}.  

The above and other advances will greatly benefit from fundamental understanding, of the underlying quantum mechanisms for transport of CDWs and other systems in which evolving topological deformations of the ground state play key roles.

\section*{Data availability}
All data generated or analysed during this study are included in this published article and its supplementary information files.

\section*{Acknowledgements}

The authors acknowledge support from the
University of Houston Grants to Enhance and Advance
Research (GEAR) program and from the State of Texas
through the Texas Center for Superconductivity at the
University of Houston.

\section*{Author contributions statement}

J.H.M. Conceptualization, methodology, formal analysis, original draft, writing review, editing, funding acquisition, and supervision. M.Y.S.V Investigation, data curation, software, formal analysis, original draft, writing review, editing,  and visualization. J.O.S. Investigation, data curation, formal analysis, and writing review. 

\section*{Competing interests}

The authors declare no competing interests.


%\appendix

%\section{Sample Appendix Section}

 \bibliographystyle{elsarticle-num} 
 \bibliography{Main}



\section*{Supplementary material}

# Supplementary Information: "Quantum Transport of Charge Density Wave Electrons in Layered Materials"

John H. Miller, Jr.,[1] Martha Y. Suárez-Villagrán,[1] and Johnathan O. Sanderson[1]

[1]*Department of Physics and Texas Center for Superconductivity, University of Houston, Houston, Tx 77204-5005 USA*

## I. BARDEEN-ZENER FORMULA PARAMETERS

Table S1. Bardeen-Zener (BZ) formula parameters used to fit the experimental results for NbS$_3$, sample 6[1] for various temperatures.

| Temperature (K) | $G_{max}(M\Omega)^{-1}$ | $V_{Tm}$ (V) | $V_0$ (V) |
|---|---|---|---|
| 297 | 9.420 | 0.087 | 0.307 |
| 346 | 5.814 | 0 | 0.063 |
| 344 | 1.680 | 0.132 | 0.452 |
| 363 | 3.746 | 0 | 0.073 |
| 373 | 1.265 | 0 | 0.457 |
| 381 | 0.917 | 0 | 0.473 |
| 398 | 0.749 | 0.064 | 0.217 |
| 417 | 0.702 | 0.00001 | 0.219 |

Table S2. BZ formula parameters used to fit experimental results for TaS$_3$[2].

| Temperature (K) | $G_{max}(M\Omega)^{-1}$ | $V_{Tm}$ (V) | $V_0$ (V) |
|---|---|---|---|
| 54 | 0.007 | 0.000265 | 7.219 |
| 73 | 0.0187 | 0.000047 | 5.192 |
| 90 | 0.05924 | 0.0000584 | 1.83531 |
| 110 | 0.05677 | 0 | 0.4756 |
| 130 | 0.3116 | 0 | 0.9878 |
| 150 | 0.15642 | 0.0973 | 0.3980 |
| 170 | 0.2637 | 0.064 | 0.62096 |
| 190 | 0.39723 | 0 | 0.715062 |
| 200 | 0.39679 | 0 | 0.6510 |
| 210 | 0.315590 | 0.00001 | 0.6215 |

Table S3. Additional BZ parameters used to fit experimental results for TaS$_3$[2]

| Temperature (K) | $G_{max}$(MΩ)$^{-1}$ | $V_{Tm}$ (V) | $V_0$ (V) |
|---|---|---|---|
| 9 | 0.0000003 | 0 | 0.221 |
| 20 | 0.000002 | 0 | 0.658 |
| 34.5 | 0.0006 | 0.001 | 12.3 |
| 42 | 0.0002 | 0.00005 | 1.913 |
| 48 | 0.0004 | 0.000029 | 1.609 |
| 54 | 0.0009 | 0.00007 | 2.088 |
| 60 | 0.0008 | 0.000078 | 1.1655 |
| 66.5 | 0.002 | 0.413 | 1.452 |
| 72.5 | 0.008 | 0.000014 | 1.083 |
| 210 | 0.316 | 0 | 0.622 |

## II. SIMULATION PARAMETERS: TUNNELING MODEL

Table S4. Scaled variables and parameters used in the simulations[3]

| Variables & Parameters | Relations |
|---|---|
| Time / relaxation time | $t' = t/\tau = t/RC$ |
| Total current | $f = \frac{\omega\tau}{2\pi} = \frac{I\tau}{Q_0} = \frac{IR}{E^*\ell}$ |
| Displacement charge | $q = \frac{\theta}{2\pi} = \frac{Q}{Q_0}$ |
| Zener / threshold field ratio | $q_0 = \frac{\theta_0}{2\pi} = \frac{E_0}{2E_T} = \frac{E_0}{E^*}$ |
| Relative displacement charge | $q'_n = \frac{\theta'_n}{2\pi} = q - n - \frac{1}{2}$ |
| Macrostate amplitudes * | $\chi_0(t) = c_0(t); \chi_1(t) = ic_1(t); c_0, c_1 \in \mathbb{R}$ |
| Relaxation parameter | $\gamma = \frac{32\pi^2 u_E \lambda \tau}{\hbar}$ |
| *Fixing $\delta$ to maximize current | |

## III. DATA USED TO GENERATE 3D COLOR PLOT FIGURE 1

| | | | | | | | | | | | | | | | | | | |
|---|---|---|---|---|---|---|---|---|---|---|---|---|---|---|---|---|---|---|
| 0.02179 | 0.05908 | 0.15775 | 0.38351 | 0.68721 | 0.87600 | 0.95315 | 0.98076 | 0.98753 | 0.98076 | 0.95315 | 0.87600 | 0.68721 | 0.38351 | 0.15775 | 0.05908 | 0.02179 | 0.00802 | 0.00295 |
| 0.04008 | 0.10804 | 0.27752 | 0.57443 | 0.81994 | 0.93205 | 0.97452 | 0.98955 | 0.99322 | 0.98955 | 0.97452 | 0.93205 | 0.81994 | 0.57443 | 0.27752 | 0.10804 | 0.04008 | 0.01476 | 0.00543 |
| 0.07137 | 0.18907 | 0.44174 | 0.73492 | 0.89730 | 0.96185 | 0.98573 | 0.99415 | 0.99621 | 0.99415 | 0.98573 | 0.96185 | 0.89730 | 0.73492 | 0.44174 | 0.18907 | 0.07137 | 0.02635 | 0.00970 |
| 0.12093 | 0.30659 | 0.60954 | 0.83869 | 0.93952 | 0.97764 | 0.99164 | 0.99657 | 0.99778 | 0.99657 | 0.99164 | 0.97764 | 0.93952 | 0.83869 | 0.60954 | 0.30659 | 0.12093 | 0.04495 | 0.01656 |
| 0.19130 | 0.44565 | 0.73786 | 0.89858 | 0.96240 | 0.98613 | 0.99481 | 0.99787 | 0.99862 | 0.99787 | 0.99481 | 0.98613 | 0.96240 | 0.89858 | 0.73786 | 0.44565 | 0.19130 | 0.07226 | 0.02668 |
| 0.27753 | 0.57445 | 0.82000 | 0.93221 | 0.97497 | 0.99077 | 0.99655 | 0.99859 | 0.99908 | 0.99859 | 0.99655 | 0.99077 | 0.97497 | 0.93221 | 0.82000 | 0.57445 | 0.27753 | 0.10805 | 0.04008 |
| 0.36542 | 0.67071 | 0.86866 | 0.95108 | 0.98197 | 0.99335 | 0.99751 | 0.99898 | 0.99934 | 0.99898 | 0.99751 | 0.99335 | 0.98197 | 0.95108 | 0.86866 | 0.67071 | 0.36542 | 0.14868 | 0.05557 |
| 0.43808 | 0.73216 | 0.89614 | 0.96150 | 0.98582 | 0.99477 | 0.99805 | 0.99920 | 0.99948 | 0.99920 | 0.99805 | 0.99477 | 0.98582 | 0.96150 | 0.89614 | 0.73216 | 0.43808 | 0.18699 | 0.07055 |
| 0.48433 | 0.76523 | 0.91010 | 0.96673 | 0.98775 | 0.99548 | 0.99831 | 0.99931 | 0.99955 | 0.99931 | 0.99831 | 0.99548 | 0.98775 | 0.96673 | 0.91010 | 0.76523 | 0.48433 | 0.21445 | 0.08157 |
| 0.50000 | 0.77558 | 0.91436 | 0.96833 | 0.98834 | 0.99570 | 0.99839 | 0.99934 | 0.99957 | 0.99934 | 0.99839 | 0.99570 | 0.98834 | 0.96833 | 0.91436 | 0.77558 | 0.50000 | 0.22442 | 0.08564 |
| 0.48433 | 0.76523 | 0.91010 | 0.96673 | 0.98775 | 0.99548 | 0.99831 | 0.99931 | 0.99955 | 0.99931 | 0.99831 | 0.99548 | 0.98775 | 0.96673 | 0.91010 | 0.76523 | 0.48433 | 0.21445 | 0.08157 |
| 0.43808 | 0.73216 | 0.89614 | 0.96150 | 0.98582 | 0.99477 | 0.99805 | 0.99920 | 0.99948 | 0.99920 | 0.99805 | 0.99477 | 0.98582 | 0.96150 | 0.89614 | 0.73216 | 0.43808 | 0.18699 | 0.07055 |
| 0.36542 | 0.67071 | 0.86866 | 0.95108 | 0.98197 | 0.99335 | 0.99751 | 0.99898 | 0.99934 | 0.99898 | 0.99751 | 0.99335 | 0.98197 | 0.95108 | 0.86866 | 0.67071 | 0.36542 | 0.14868 | 0.05557 |
| 0.27753 | 0.57445 | 0.82000 | 0.93221 | 0.97497 | 0.99077 | 0.99655 | 0.99859 | 0.99908 | 0.99859 | 0.99655 | 0.99077 | 0.97497 | 0.93221 | 0.82000 | 0.57445 | 0.27753 | 0.10805 | 0.04008 |
| 0.19130 | 0.44565 | 0.73786 | 0.89858 | 0.96240 | 0.98613 | 0.99481 | 0.99787 | 0.99862 | 0.99787 | 0.99481 | 0.98613 | 0.96240 | 0.89858 | 0.73786 | 0.44565 | 0.19130 | 0.07226 | 0.02668 |
| 0.12093 | 0.30659 | 0.60954 | 0.83869 | 0.93952 | 0.97764 | 0.99164 | 0.99657 | 0.99778 | 0.99657 | 0.99164 | 0.97764 | 0.93952 | 0.83869 | 0.60954 | 0.30659 | 0.12093 | 0.04495 | 0.01656 |
| 0.07137 | 0.18907 | 0.44174 | 0.73492 | 0.89730 | 0.96185 | 0.98573 | 0.99415 | 0.99621 | 0.99415 | 0.98573 | 0.96185 | 0.89730 | 0.73492 | 0.44174 | 0.18907 | 0.07137 | 0.02635 | 0.00970 |
| 0.04008 | 0.10804 | 0.27752 | 0.57443 | 0.81994 | 0.93205 | 0.97452 | 0.98955 | 0.99322 | 0.98955 | 0.97452 | 0.93205 | 0.81994 | 0.57443 | 0.27752 | 0.10804 | 0.04008 | 0.01476 | 0.00543 |
| 0.02179 | 0.05908 | 0.15775 | 0.38351 | 0.68721 | 0.87600 | 0.95315 | 0.98076 | 0.98753 | 0.98076 | 0.95315 | 0.87600 | 0.68721 | 0.38351 | 0.15775 | 0.05908 | 0.02179 | 0.00802 | 0.00295 |
| 0.01165 | 0.03166 | 0.08561 | 0.22434 | 0.49979 | 0.77500 | 0.91279 | 0.96404 | 0.97668 | 0.96404 | 0.91279 | 0.77500 | 0.49979 | 0.22434 | 0.08561 | 0.03166 | 0.01165 | 0.00429 | 0.00158 |

# IV. FIGURE 2: RAW DATA OF QUANTUM THEORETICAL BARDEEN-ZENER (SOLID LINES) $I-V$ CURVES FOR $NbS_3$ AT VARIOUS TEMPERATURES RANGING FROM ROOM TEMPERATURE UP TO 474 K.

| T=293 K | |
|---|---|
| Voltage (V) | CDW current (µA) |
| 0.26999259 | 0.001490067 |
| 0.30070995 | 0.0034751727 |
| 0.33582456 | 0.0069583241 |
| 0.36215312 | 0.010496159 |
| 0.39505432 | 0.016082287 |
| 0.4257928 | 0.022484681 |
| 0.46529199 | 0.032372574 |
| 0.49600935 | 0.041318008 |
| 0.52891477 | 0.052070941 |
| 0.55305527 | 0.060697744 |
| 0.57499504 | 0.069057628 |
| 0.60570817 | 0.081555849 |
| 0.64081011 | 0.096917934 |
| 0.67809588 | 0.11441648 |
| 0.71318937 | 0.13192303 |
| 0.74828708 | 0.15036888 |
| 0.78338056 | 0.16968653 |
| 0.8140937 | 0.18726362 |
| 0.84480261 | 0.20542536 |
| 0.88209261 | 0.22821875 |
| 0.91499381 | 0.24895993 |
| 0.95008307 | 0.27168867 |
| 0.98078354 | 0.29205786 |
| 1.0114882 | 0.31285411 |
| 1.0465859 | 0.3371149 |
| 1.0794787 | 0.36029713 |
| 1.1123799 | 0.38389078 |

| T=321 K | |
|---|---|
| Voltage (V) | CDW current (µA) |
| 0.27027924 | 0.0042840752 |
| 0.29926685 | 0.0099732137 |
| 0.32825739 | 0.017777385 |
| 0.35573138 | 0.027141745 |
| 0.39082334 | 0.041832689 |
| 0.43199032 | 0.062792708 |
| 0.45792429 | 0.077939699 |
| 0.49450347 | 0.10167817 |
| 0.52195399 | 0.12119634 |
| 0.54482845 | 0.13850184 |
| 0.58292125 | 0.16927418 |
| 0.61798681 | 0.19959325 |
| 0.65456306 | 0.23307432 |
| 0.69419295 | 0.27129903 |
| 0.72923504 | 0.30663732 |
| 0.76275178 | 0.34167323 |
| 0.78865348 | 0.3695154 |
| 0.82368971 | 0.40816218 |
| 0.85566936 | 0.44435702 |
| 0.87852622 | 0.47072806 |

| T=344 K | |
|---|---|
| Voltage (V) | CDW current (µA) |
| 0.10613053 | 0.00060668786 |
| 0.13813071 | 0.00041599154 |
| 0.17324614 | 0.005155823 |
| 0.20226308 | 0.012718302 |
| 0.23739318 | 0.026500736 |
| 0.26795308 | 0.042435586 |
| 0.30155195 | 0.063823095 |
| 0.33206198 | 0.086390403 |
| 0.35954183 | 0.10899763 |
| 0.39461033 | 0.14062454 |
| 0.41901307 | 0.16426724 |
| 0.44645479 | 0.19228065 |
| 0.47389065 | 0.22164744 |
| 0.49218024 | 0.24191041 |
| 0.51656831 | 0.26971035 |
| 0.54703728 | 0.30558551 |
| 0.57446433 | 0.3388609 |
| 0.60950642 | 0.38258598 |
| 0.64606801 | 0.42949768 |
| 0.67805646 | 0.47150304 |
| 0.71614339 | 0.5225578 |
| 0.745087 | 0.56203743 |
| 0.77554716 | 0.60416025 |
| 0.81210581 | 0.65542355 |
| 0.8517093 | 0.71174136 |
| 0.88064411 | 0.75335522 |
| 0.90501752 | 0.78869005 |
| 0.93243578 | 0.82872493 |
| 0.96442129 | 0.8757843 |
| 0.9979263 | 0.92545864 |
| 1.0314372 | 0.9754993 |
| 1.0558018 | 1.0120906 |
| 1.0755962 | 1.0419398 |
| 1.1014774 | 1.0811231 |

| T=370 K | |
|---|---|
| Voltage (V) | CDW current (µA) |
| 0.16941515 | 0.0093389142 |
| 0.20157667 | 0.023905212 |
| 0.23063175 | 0.041767866 |
| 0.2626906 | 0.066006774 |
| 0.29476705 | 0.094334528 |
| 0.32376346 | 0.12292609 |
| 0.35426763 | 0.15561046 |
| 0.38933612 | 0.19599294 |
| 0.41982562 | 0.23318039 |
| 0.45488531 | 0.27797182 |
| 0.48079874 | 0.31228769 |
| 0.51585257 | 0.3601236 |
| 0.54937811 | 0.40720301 |
| 0.58594262 | 0.45982997 |
| 0.61182966 | 0.49779881 |
| 0.64840297 | 0.55232442 |
| 0.67429587 | 0.59148731 |
| 0.70628725 | 0.64044674 |

| T=415 K | |
|---|---|
| Voltage (V) | CDW current (µA) |
| 0.05140549 | 0.00066890249 |
| 0.074271146 | 0.00070276477 |
| 0.097195469 | 0.0028862594 |
| 0.1417417 | 0.024490305 |
| 0.17244826 | 0.049941876 |
| 0.20303749 | 0.082205147 |
| 0.23057015 | 0.11599199 |
| 0.25653052 | 0.15119094 |
| 0.27943137 | 0.18447547 |
| 0.29926098 | 0.21473338 |
| 0.32060128 | 0.24858353 |
| 0.34042796 | 0.28107784 |
| 0.35872929 | 0.31185903 |
| 0.38159788 | 0.35125833 |
| 0.40141282 | 0.38614175 |
| 0.42275606 | 0.42439987 |
| 0.44561585 | 0.46607553 |
| 0.47152928 | 0.51408993 |
| 0.49133249 | 0.55127391 |
| 0.51723126 | 0.60047335 |
| 0.54312709 | 0.65024337 |
| 0.57055121 | 0.70350655 |
| 0.588832 | 0.73929696 |
| 0.61321128 | 0.78734819 |
| 0.64366264 | 0.84783327 |
| 0.67564229 | 0.91184975 |
| 0.70915023 | 0.97940686 |
| 0.74113868 | 1.0443075 |
| 0.77313006 | 1.1095679 |
| 0.80511265 | 1.1751269 |
| 0.83404159 | 1.2346723 |

| T=433 K | |
|---|---|
| Voltage (V) | CDW current (µA) |
| 0.1209734 | 0.0099788562 |
| 0.15485388 | 0.037154854 |
| 0.18243054 | 0.066975732 |
| 0.21602941 | 0.11033786 |
| 0.23131229 | 0.13209202 |
| 0.24658637 | 0.15488518 |
| 0.26490237 | 0.1834282 |
| 0.28321542 | 0.21311495 |
| 0.30001779 | 0.24123199 |
| 0.3168055 | 0.27006239 |
| 0.33663218 | 0.30494723 |
| 0.35797248 | 0.3433795 |
| 0.376265 | 0.37696195 |
| 0.39913066 | 0.41966565 |
| 0.41741145 | 0.45431984 |
| 0.43569517 | 0.48938479 |

| T=454 K | |
|---|---|
| Voltage (V) | CDW current (µA) |
| 0.094473297 | 0.0010944466 |
| 0.13465466 | 0.034358186 |
| 0.16838554 | 0.075200359 |
| 0.20053239 | 0.1213425 |
| 0.22343912 | 0.15730908 |
| 0.26769201 | 0.23200941 |
| 0.29055766 | 0.27264437 |
| 0.30885019 | 0.30592953 |
| 0.34389521 | 0.37127435 |
| 0.3758778 | 0.43237218 |
| 0.40481261 | 0.48860754 |
| 0.43984003 | 0.55767377 |
| 0.47334504 | 0.62456965 |

| T=470 K | |
|---|---|
| Voltage (V) | CDW current (µA) |
| 0.12167741 | 0.0090959622 |
| 0.15703925 | 0.030687203 |
| 0.18307588 | 0.053343905 |
| 0.21677449 | 0.089622596 |
| 0.24122123 | 0.11996865 |
| 0.26106845 | 0.14667064 |
| 0.27480837 | 0.16610127 |
| 0.2839693 | 0.17944556 |
| 0.30075114 | 0.20462926 |
| 0.31297158 | 0.22352029 |
| 0.33585483 | 0.26001369 |
| 0.36329655 | 0.30546576 |
| 0.3876817 | 0.34718364 |
| 0.41969068 | 0.40355728 |
| 0.45321329 | 0.46425974 |
| 0.49130316 | 0.53495415 |

| T=474 K | |
|---|---|
| Voltage (V) | CDW current (µA) |
| 0.034036729 | 4.47E-08 |
| 0.055072296 | 1.48E-05 |
| 0.072274503 | 0.00015033315 |
| 0.095217767 | 0.00096178936 |
| 0.11624435 | 0.0028892632 |
| 0.13152688 | 0.0052503973 |
| 0.14680342 | 0.0085266847 |
| 0.16972872 | 0.015247533 |
| 0.18692194 | 0.021710109 |
| 0.20028779 | 0.027551676 |
| 0.21174595 | 0.033107391 |
| 0.22702848 | 0.041269788 |
| 0.24804308 | 0.053814388 |
| 0.26331363 | 0.063822548 |
| 0.27285211 | 0.070430027 |
| 0.28430728 | 0.078707544 |
| 0.30341718 | 0.093297697 |
| 0.32443777 | 0.11038738 |
| 0.34736606 | 0.13015345 |
| 0.36265159 | 0.14392494 |
| 0.38558287 | 0.16539498 |
| 0.41043383 | 0.18965871 |
| 0.43145742 | 0.21091456 |
| 0.4543917 | 0.2347902 |
| 0.47542128 | 0.25725707 |
| 0.49357883 | 0.27706003 |
| 0.51078553 | 0.29614568 |
| 0.52895057 | 0.3166081 |
| 0.54615876 | 0.33626972 |
| 0.57101571 | 0.36511043 |

## V. FIGURE 3A: RAW DATA OF LN(G) VS. $V_0/V$ FOR $NbS_3$ S#4

| T=346 K | |
|---|---|
| ln(g) | V0/V |
| 0.3966818 | -0.38252046 |
| 0.3514902 | -0.34873213 |
| 0.31199692 | -0.31786939 |
| 0.27223261 | -0.28500812 |
| 0.24357654 | -0.25978871 |
| 0.21525369 | -0.23334058 |
| 0.1983409 | -0.21653318 |
| 0.18030991 | -0.19773579 |
| 0.16528408 | -0.18149755 |
| 0.15009582 | -0.16418963 |
| 0.1361163 | -0.14736573 |
| 0.12679327 | -0.13570211 |
| 0.11618295 | -0.12186972 |
| 0.10804563 | -0.11073233 |
| 0.10246393 | -0.10290163 |
| 0.09575078 | -0.093399362 |
| 0.090154955 | -0.085206994 |
| 0.084657701 | -0.076721127 |
| 0.080022265 | -0.069352352 |

| T=363 K | |
|---|---|
| ln(g) | V0/V |
| 0.48440565 | -0.46408133 |
| 0.4305828 | -0.42927367 |
| 0.37502373 | -0.38869179 |
| 0.33861366 | -0.35916798 |
| 0.30594041 | -0.33052928 |
| 0.28824138 | -0.31384984 |
| 0.26623822 | -0.29204069 |
| 0.24220283 | -0.2670711 |
| 0.21662862 | -0.23862653 |
| 0.19484473 | -0.2128071 |
| 0.18165212 | -0.19641435 |
| 0.16608194 | -0.17643739 |
| 0.14904789 | -0.15361232 |
| 0.1406339 | -0.14175642 |
| 0.1316121 | -0.12859325 |
| 0.12545758 | -0.11927501 |
| 0.11862996 | -0.10867888 |

| T=368 K | |
|---|---|
| ln(g) | V0/V |
| 0.49914412 | -0.51316309 |
| 0.46163618 | -0.47768357 |
| 0.42707518 | -0.44357016 |
| 0.3993153 | -0.41523099 |
| 0.37671254 | -0.39168947 |
| 0.35027658 | -0.36339985 |
| 0.33001264 | -0.3412279 |
| 0.3107512 | -0.31973884 |
| 0.29147102 | -0.2979361 |
| 0.2782685 | -0.28284695 |
| 0.26099039 | -0.2627174 |
| 0.24725405 | -0.24627764 |
| 0.23558425 | -0.23209565 |
| 0.22560186 | -0.21993361 |
| 0.21584611 | -0.20799899 |
| 0.21072047 | -0.2017334 |

| T=373 K | |
|---|---|
| ln(g) | V0/V |
| 1.0503175 | -1.0346766 |
| 0.96029027 | -0.95296612 |
| 0.88447788 | -0.88428173 |
| 0.83676745 | -0.84072293 |
| 0.76386726 | -0.77166237 |
| 0.72539912 | -0.73447616 |
| 0.68359646 | -0.69395604 |
| 0.63615443 | -0.64661749 |
| 0.60377532 | -0.61324003 |
| 0.57290044 | -0.58113511 |
| 0.53208695 | -0.53850472 |
| 0.50541593 | -0.51013236 |
| 0.47673985 | -0.47857096 |
| 0.44615256 | -0.4431617 |
| 0.42998072 | -0.4239636 |
| 0.41408821 | -0.4046722 |

| T=381 K | |
|---|---|
| ln(g) | V0/V |
| 1.027725 | -1.0230638 |
| 0.95264003 | -0.94934807 |
| 0.88401765 | -0.88304308 |
| 0.81814967 | -0.81918475 |
| 0.76420574 | -0.76632262 |
| 0.70721412 | -0.71040928 |
| 0.66654366 | -0.67015814 |
| 0.62091716 | -0.6256104 |
| 0.57791736 | -0.58356473 |
| 0.54614703 | -0.5512847 |
| 0.51386248 | -0.51711305 |
| 0.48973748 | -0.49120413 |
| 0.46882734 | -0.4682873 |
| 0.44769993 | -0.4442919 |
| 0.42927606 | -0.42382693 |

| T=398 K | |
|---|---|
| ln(g) | V0/V |
| 1.0745141 | -1.0992652 |
| 0.97583424 | -0.99331814 |
| 0.89375472 | -0.90179596 |
| 0.81736543 | -0.81453124 |
| 0.75898218 | -0.74990548 |
| 0.70317467 | -0.69200214 |
| 0.65055616 | -0.64074395 |
| 0.59769847 | -0.59145756 |
| 0.54646717 | -0.54387053 |
| 0.50332503 | -0.50307144 |
| 0.46423182 | -0.46591501 |
| 0.4346898 | -0.43733554 |
| 0.4018141 | -0.40480805 |
| 0.37502649 | -0.37762136 |
| 0.34712071 | -0.34741934 |
| 0.32750601 | -0.32498736 |

| T=417 K | |
|---|---|
| ln(g) | V0/V |
| 1.5726267 | -1.5117769 |
| 1.2987135 | -1.2635001 |
| 1.0881296 | -1.0700057 |
| 0.93630882 | -0.93251278 |
| 0.81172738 | -0.81579328 |
| 0.72413594 | -0.73247207 |
| 0.66003547 | -0.67019927 |
| 0.60093064 | -0.60961399 |
| 0.55154124 | -0.55701571 |
| 0.50202829 | -0.50363973 |
| 0.47035947 | -0.47027272 |
| 0.44244901 | -0.44176316 |
| 0.42028111 | -0.41955111 |
| 0.40022858 | -0.39939832 |
| 0.37770232 | -0.37609886 |
| 0.35949219 | -0.35701786 |

# VI. FIGURE 3A: RAW DATA OF LN(G) VS. $V_0/V$ FOR $NbS_3$ S#6

| T=293 K | |
|---|---|
| ln(g) | V0/V |
| 2.5999872 | -2.6362883 |
| 2.3834532 | -2.4123109 |
| 2.2113889 | -2.2330348 |
| 2.0236615 | -2.0253382 |
| 1.8983382 | -1.8920924 |
| 1.7802367 | -1.7726561 |
| 1.7025305 | -1.694786 |
| 1.637568 | -1.6287734 |
| 1.5545332 | -1.5453728 |
| 1.4693799 | -1.4604384 |
| 1.3885846 | -1.3814466 |
| 1.3202573 | -1.3165454 |
| 1.2583319 | -1.2578405 |
| 1.2019618 | -1.2041285 |
| 1.1566156 | -1.160258 |
| 1.1145722 | -1.1186626 |
| 1.0674542 | -1.0712608 |
| 1.0290709 | -1.0319923 |
| 0.99106436 | -0.99299057 |
| 0.96004209 | -0.96146 |
| 0.93089909 | -0.93190632 |
| 0.89968099 | -0.89949035 |
| 0.87226685 | -0.87061919 |
| 0.84646755 | -0.84321496 |

| T=321 K | |
|---|---|
| ln(g) | V0/V |
| 2.365686 | -2.4097635 |
| 2.1567569 | -2.2082942 |
| 1.9901854 | -2.0257696 |
| 1.811487 | -1.823165 |
| 1.6388594 | -1.6399276 |
| 1.5460447 | -1.5442947 |
| 1.4316814 | -1.4294332 |
| 1.3563866 | -1.3553332 |
| 1.2994391 | -1.297728 |
| 1.2145232 | -1.2135255 |
| 1.1456092 | -1.1458947 |
| 1.0815939 | -1.0820433 |
| 1.0198482 | -1.019715 |
| 0.97084117 | -0.97001145 |
| 0.9281806 | -0.92733101 |
| 0.89769642 | -0.89694877 |
| 0.85951226 | -0.85905454 |
| 0.82738899 | -0.82787621 |
| 0.80586258 | -0.80709384 |

| T=344 K | |
|---|---|
| ln(g) | V0/V |
| 2.2334202 | -2.2705288 |
| 1.9029126 | -1.9558523 |
| 1.6858865 | -1.7272886 |
| 1.4980452 | -1.5194993 |
| 1.3604041 | -1.3712597 |
| 1.2564281 | -1.2603902 |
| 1.144771 | -1.1426117 |
| 1.0781011 | -1.0739722 |
| 1.0118347 | -1.0060985 |
| 0.95325466 | -0.94696079 |
| 0.91783138 | -0.91155941 |
| 0.87449898 | -0.8682993 |
| 0.82579101 | -0.82026552 |
| 0.78636469 | -0.78196875 |
| 0.74115456 | -0.73818719 |
| 0.69921194 | -0.69773557 |
| 0.66622544 | -0.66607767 |
| 0.63079331 | -0.63189966 |
| 0.60628955 | -0.60796522 |
| 0.5824771 | -0.58467863 |
| 0.55625567 | -0.5589281 |
| 0.53039043 | -0.53313997 |
| 0.5129637 | -0.51566235 |
| 0.49914886 | -0.50170174 |
| 0.4844714 | -0.48660088 |
| 0.46840366 | -0.46981077 |
| 0.45267718 | -0.45320679 |
| 0.43796993 | -0.4374364 |
| 0.42786295 | -0.42646922 |
| 0.4199889 | -0.41795074 |
| 0.41012052 | -0.40739166 |

| T=370 K | |
|---|---|
| ln(g) | V0/V |
| 1.9447436 | -2.0297466 |
| 1.6344601 | -1.6994139 |
| 1.4285502 | -1.4525066 |
| 1.2542094 | -1.2623976 |
| 1.1177268 | -1.1201175 |
| 1.0176227 | -1.0151035 |
| 0.9300004 | -0.9254617 |
| 0.8462329 | -0.84229696 |
| 0.78477592 | -0.78234168 |
| 0.72429034 | -0.72358315 |
| 0.68525353 | -0.6853583 |
| 0.63868836 | -0.63922088 |
| 0.59971271 | -0.60013913 |
| 0.5622889 | -0.56244188 |
| 0.53849798 | -0.53884988 |
| 0.50812388 | -0.508578 |
| 0.48861197 | -0.48876731 |
| 0.46648022 | -0.46618514 |

| T=415 K | |
|---|---|
| ln(g) | V0/V |
| 1.3727687 | -1.4156484 |
| 1.16595 | -1.1799659 |
| 1.0267226 | -1.0295244 |
| 0.92282032 | -0.91929136 |
| 0.84719039 | -0.84099905 |
| 0.79105391 | -0.78395867 |
| 0.7383987 | -0.73226894 |
| 0.69539402 | -0.69092839 |
| 0.65991704 | -0.65669955 |
| 0.6203692 | -0.61840927 |
| 0.58974591 | -0.58878452 |
| 0.55997204 | -0.55982958 |
| 0.53124586 | -0.53168632 |
| 0.50205063 | -0.50275294 |
| 0.48181542 | -0.48257279 |
| 0.45769 | -0.45865274 |
| 0.43586773 | -0.43707222 |
| 0.41491731 | -0.41625655 |
| 0.40203585 | -0.40333205 |
| 0.38605222 | -0.38706896 |
| 0.36778828 | -0.36851224 |
| 0.35038004 | -0.3511176 |
| 0.33382429 | -0.33455274 |
| 0.31941603 | -0.31989658 |
| 0.3061989 | -0.3060796 |
| 0.29403534 | -0.29309374 |
| 0.28383665 | -0.28220624 |

| T=433 K | |
|---|---|
| ln(g) | V0/V |
| 1.4961369 | -1.4957756 |
| 1.1687971 | -1.1950833 |
| 0.99211881 | -0.98707553 |
| 0.83781541 | -0.83030269 |
| 0.78246066 | -0.77779344 |
| 0.73399339 | -0.73139174 |
| 0.68324331 | -0.68291541 |
| 0.63906395 | -0.64082629 |
| 0.60327344 | -0.60572746 |
| 0.57130564 | -0.57321252 |
| 0.53765736 | -0.53864167 |
| 0.50560526 | -0.50595204 |
| 0.48102472 | -0.48108778 |
| 0.45346747 | -0.45314295 |
| 0.43360758 | -0.43310611 |
| 0.41541146 | -0.41495929 |

| T=454 K | |
|---|---|
| ln(g) | V0/V |
| 0.6574538 | -0.67429719 |
| 0.59005239 | -0.59660368 |
| 0.49250923 | -0.49016943 |
| 0.45375084 | -0.44923888 |
| 0.42687617 | -0.4215931 |
| 0.38337487 | -0.37810474 |
| 0.35075438 | -0.34695492 |
| 0.32568349 | -0.32408905 |
| 0.29974712 | -0.30118115 |
| 0.27852998 | -0.28276389 |

| T=470 K | |
|---|---|
| ln(g) | V0/V |
| 1.1425457 | -1.1356357 |
| 0.96493166 | -0.96759034 |
| 0.86713995 | -0.8673663 |
| 0.80121734 | -0.8013441 |
| 0.76115791 | -0.76167797 |
| 0.73660274 | -0.73727803 |
| 0.6955005 | -0.69656072 |
| 0.66834365 | -0.66927306 |
| 0.62280648 | -0.62259511 |
| 0.5757626 | -0.57476213 |
| 0.53954718 | -0.53862294 |
| 0.49839698 | -0.49790943 |
| 0.46153229 | -0.46150906 |
| 0.4257505 | -0.42634182 |

| T=474 K | |
|---|---|
| ln(g) | V0/V |
| 1.9981163 | -1.9422928 |
| 1.890422 | -1.8579237 |
| 1.7808083 | -1.7659982 |
| 1.6381949 | -1.6387978 |
| 1.5073989 | -1.5180583 |
| 1.3880367 | -1.4044515 |
| 1.2662501 | -1.285848 |
| 1.2041205 | -1.2237964 |
| 1.1478025 | -1.165781 |
| 1.077249 | -1.0909512 |
| 1.0190746 | -1.0277589 |
| 0.95183768 | -0.95375026 |
| 0.90948014 | -0.90671478 |
| 0.86458701 | -0.85649249 |
| 0.83800223 | -0.82639986 |

# VII. FIGURE 3B: RAW DATA OF LN(G) VS. $V_0/V$ FOR NbSe$_3$

| T=70 K | | T=86 K | | T=99 K | | T=114 K | | T=125 K | |
|---|---|---|---|---|---|---|---|---|---|
| ln(g) | V0/V | ln(g) | V0/V | ln(g) | V0/V | ln(g) | V0/V | ln(g) | V0/V |
| 1.5199372 | -1.5102187 | 0.98948984 | -0.96121931 | 1.5933483 | -1.5903469 | 1.4938756 | -1.4862664 | 1.4705712 | -1.5109319 |
| 1.4362946 | -1.4086126 | 0.92625853 | -0.91063781 | 1.5056659 | -1.4890547 | 1.3863718 | -1.3806665 | 1.3755156 | -1.3806665 |
| 1.3131727 | -1.3270134 | 0.83890646 | -0.85136255 | 1.3765974 | -1.3706948 | 1.2665333 | -1.2710284 | 1.3121408 | -1.3081955 |
| 1.2177152 | -1.244989 | 0.69139253 | -0.69587198 | 1.2825648 | -1.2417131 | 1.1893322 | -1.1901898 | 1.2516859 | -1.2406235 |
| 1.1781651 | -1.1631206 | 0.64416489 | -0.64058485 | 1.1291967 | -1.1158807 | 1.1037443 | -1.0893207 | 1.1987189 | -1.1901898 |
| 1.0620353 | -1.0570831 | 0.60871632 | -0.60362573 | 1.0771706 | -1.0735326 | 0.98483354 | -0.98708038 | 1.1168368 | -1.1178051 |
| 0.98483354 | -0.98877203 | 0.4630218 | -0.48945076 | 0.94836335 | -0.95138543 | 0.92117532 | -0.92165911 | 1.0737887 | -1.0707722 |
| 0.77792441 | -0.76652282 | 0.41738251 | -0.41807259 | 0.83495876 | -0.83816858 | 0.84819047 | -0.86585523 | 1.0202971 | -1.0302545 |
| 0.73165574 | -0.7286138 | 0.33597182 | -0.35414064 | 0.69139253 | -0.69587198 | 0.78715212 | -0.78672165 | 0.97328843 | -0.98286374 |
| 0.6979458 | -0.69208954 | 0.28890202 | -0.31583999 | 0.64416489 | -0.64417707 | 0.72478595 | -0.72959321 | 0.92117532 | -0.91770897 |
| 0.65953765 | -0.66233429 | 0.24263629 | -0.24458301 | 0.60300086 | -0.61230242 | 0.66998944 | -0.681761 | 0.87185251 | -0.8828428 |
| 0.62618981 | -0.63522061 | 0.20766101 | -0.21369289 | 0.53340453 | -0.54328167 | 0.57930277 | -0.58819491 | 0.78098823 | -0.79192267 |
| 0.60016328 | -0.60535506 | 0.17195492 | -0.17693924 | 0.47184078 | -0.47871874 | 0.52508346 | -0.52767972 | 0.72764049 | -0.73614726 |
| 0.55653602 | -0.55144818 | 0.12956989 | -0.12958712 | 0.43343335 | -0.44867563 | 0.45759384 | -0.4668441 | 0.66736106 | -0.67555281 |
| 0.51123447 | -0.5255467 | 0.1167984 | -0.11041231 | 0.38522069 | -0.40382247 | 0.42633652 | -0.43593211 | 0.6072824 | -0.61988361 |
| 0.4876801 | -0.49563546 | 0.097172878 | -0.093671312 | 0.34889193 | -0.36089504 | 0.37891328 | -0.38605901 | 0.56358052 | -0.57132428 |
| 0.45651591 | -0.46208345 | 0.083165706 | -0.078227148 | 0.31008318 | -0.31973479 | 0.32762533 | -0.33634879 | 0.51284461 | -0.51838118 |
| 0.43343335 | -0.4281699 | 0.069191452 | -0.067051123 | 0.24263629 | -0.25063439 | 0.31873359 | -0.32212243 | 0.44868827 | -0.45065067 |
| 0.41541841 | -0.41092215 | 0.045044125 | -0.046055811 | 0.20282077 | -0.20786033 | 0.28439517 | -0.29215161 | 0.38340793 | -0.39303326 |
| 0.35890704 | -0.35010978 | 0.031178447 | -0.030350121 | 0.15870477 | -0.16907633 | 0.25777746 | -0.26305296 | 0.31998891 | -0.32757018 |
| 0.31008318 | -0.30809562 | 0.021992005 | -0.023555693 | 0.13906945 | -0.14257769 | 0.2228859 | -0.23179449 | 0.25981195 | -0.2692175 |
| 0.27300366 | -0.27895234 | | | 0.085553018 | -0.093671312 | 0.18311675 | -0.19189864 | 0.23457125 | -0.23976813 |
| 0.25078139 | -0.25306523 | | | 0.076036587 | -0.081296958 | 0.14636051 | -0.16275609 | 0.21261677 | -0.21408293 |
| 0.23145718 | -0.23378195 | | | 0.048575166 | -0.053005385 | 0.12654983 | -0.14076333 | 0.17884859 | -0.17769133 |
| 0.18456198 | -0.18486447 | | | 0.037652587 | -0.042106203 | 0.087801539 | -0.10515058 | 0.14751565 | -0.14804064 |
| 0.15795794 | -0.16127476 | | | 0.028371631 | -0.030350121 | 0.072704582 | -0.082664349 | 0.10942063 | -0.10777798 |
| 0.12015116 | -0.12529417 | | | | | 0.054999451 | -0.063185611 | 0.088147342 | -0.087808781 |
| 0.098557708 | -0.11252478 | | | | | 0.049656259 | -0.055665598 | 0.072704582 | -0.071608047 |
| 0.080845012 | -0.083348746 | | | | | 0.038159171 | -0.040793125 | 0.062371365 | -0.056498368 |
| 0.069518589 | -0.067051123 | | | | | 0.028640548 | -0.032625227 | 0.052672066 | -0.04655061 |
| 0.050921288 | -0.053005385 | | | | | | | 0.045902061 | -0.036700837 |
| 0.042766583 | -0.044079057 | | | | | | | 0.0412801 | -0.030187811 |
| 0.034101709 | -0.036210886 | | | | | | | 0.037416527 | -0.026947136 |
| 0.031178447 | -0.031324533 | | | | | | | 0.030982976 | -0.016486891 |
| 0.027320977 | -0.027432568 | | | | | | | 0.026684169 | -0.013290243 |
| 0.024627994 | -0.024523504 | | | | | | | 0.023163193 | -0.010103782 |
| | | | | | | | | 0.02050589 | -0.010899446 |

# VIII. FIGURE 3C: RAW DATA OF LN(G) VS. $V_0/V$ FOR TaS$_3$

| T=9 K | |
|---|---|
| ln(g) | V0/V |
| 1.6986677 | -1.6270868 |
| 1.2387914 | -1.2441543 |
| 0.97014908 | -0.98810836 |
| 0.73233401 | -0.74360392 |
| 0.58878447 | -0.58165733 |

| T=20 K | |
|---|---|
| ln(g) | V0/V |
| 1.157416 | -1.1654467 |
| 0.98434482 | -1.0183471 |
| 0.84082612 | -0.87891466 |
| 0.74383097 | -0.77140205 |
| 0.66964847 | -0.67624969 |
| 0.60023092 | -0.56816234 |

| T=35 K | |
|---|---|
| ln(g) | V0/V |
| 2.2883919 | -2.1888406 |
| 1.8957512 | -1.9243315 |
| 1.5842896 | -1.6361996 |
| 1.3832453 | -1.4114825 |
| 1.3239995 | -1.3385732 |
| 1.2024381 | -1.1702376 |

| T=42 K | |
|---|---|
| ln(g) | V0/V |
| 1.2493926 | -1.2946834 |
| 0.92367718 | -0.99130931 |
| 0.60941319 | -0.59634927 |

| T=48 K | |
|---|---|
| ln(g) | V0/V |
| 0.77460492 | -0.85680757 |
| 0.50439232 | -0.55496793 |
| 0.36482455 | -0.34165017 |

| T=54 K | |
|---|---|
| ln(g) | V0/V |
| 1.0097281 | -0.99302276 |
| 0.65176401 | -0.72365167 |
| 0.48396373 | -0.51555421 |
| 0.38712805 | -0.36086704 |

| T=60 K | |
|---|---|
| ln(g) | V0/V |
| 1.7511383 | -1.7234794 |
| 1.4570436 | -1.4714114 |
| 1.2721467 | -1.2896843 |
| 1.1452751 | -1.1616455 |
| 0.75893151 | -0.77428677 |
| 0.55375856 | -0.54763226 |

| T=67 K | |
|---|---|
| ln(g) | V0/V |
| 2.9760957 | -2.8001428 |
| 2.4015465 | -2.6054161 |
| 2.078516 | -2.0604427 |
| 1.814755 | -1.7550627 |
| 1.6124535 | -1.5995899 |
| 1.4140116 | -1.4367646 |
| 0.91672707 | -0.91596691 |

| T=73 K | |
|---|---|
| ln(g) | V0/V |
| 2.8486575 | -2.8135966 |
| 2.248944 | -2.2239537 |
| 1.854934 | -1.84599 |
| 1.5569792 | -1.5667378 |
| 1.3126179 | -1.3231764 |
| 1.1765485 | -1.1695108 |

## IX. FIGURE 3C: RAW DATA OF LN(G) VS. $V_0/V$ FOR TaS$_3$

| T=54 K | |
|---|---|
| ln(g) | V0/V |
| 2.4623401 | -2.3944225 |
| 2.1475847 | -2.2014121 |
| 1.6020255 | -1.6853413 |
| 1.4434908 | -1.3986936 |

| T=73 K | |
|---|---|
| ln(g) | V0/V |
| 2.9433526 | -2.7653093 |
| 2.1601816 | -2.2514533 |
| 1.6539533 | -1.6413249 |

| T=90 K | |
|---|---|
| ln(g) | V0/V |
| 2.2731074 | -2.072513 |
| 1.2006889 | -1.2590151 |
| 0.92231171 | -0.90663984 |

| T=110 K | |
|---|---|
| ln(g) | V0/V |
| 5.0531149 | -4.9514643 |
| 4.4941423 | -4.4024022 |
| 3.1616208 | -3.0917687 |
| 2.5008352 | -2.4759337 |
| 1.7882128 | -1.8026393 |
| 1.3252867 | -1.3231558 |

| T=130 K | |
|---|---|
| ln(g) | V0/V |
| 1.146319 | -1.1788193 |
| 0.70794441 | -0.75936754 |
| 0.60945899 | -0.62992754 |
| 0.51452257 | -0.4866855 |

| T=150 K | |
|---|---|
| ln(g) | V0/V |
| 3.7122804 | -3.5578745 |
| 3.2801973 | -3.3934258 |
| 2.8423252 | -2.8886676 |
| 2.5777856 | -2.5725403 |
| 2.2629634 | -2.256333 |
| 2.1621124 | -2.1610004 |
| 1.8980559 | -1.8956768 |
| 1.6234012 | -1.6252097 |
| 1.4158855 | -1.4155949 |

| T=170 K | |
|---|---|
| ln(g) | V0/V |
| 4.3770208 | -4.2224954 |
| 4.0216899 | -3.9537052 |
| 3.6119267 | -3.6208911 |
| 3.181148 | -3.2222929 |
| 2.9515935 | -2.979485 |
| 2.6165556 | -2.621096 |
| 2.2672727 | -2.2606076 |

| T=190 K | |
|---|---|
| ln(g) | V0/V |
| 2.7684358 | -2.5470003 |
| 1.7491528 | -1.7280491 |
| 1.4670984 | -1.4761719 |
| 1.179521 | -1.1954554 |
| 0.99578512 | -0.99890808 |
| 0.86849614 | -0.86270912 |

| T=200 K | |
|---|---|
| ln(g) | V0/V |
| 2.0868072 | -2.0228578 |
| 1.6723014 | -1.6636919 |
| 1.3621281 | -1.3769361 |
| 1.070446 | -1.082678 |
| 0.92453711 | -0.92759392 |
| 0.78818285 | -0.78273161 |

| T=210 K | |
|---|---|
| ln(g) | V0/V |
| 2.5024794 | -2.4156728 |
| 1.9923897 | -1.9656177 |
| 1.6820319 | -1.6811691 |
| 1.2962706 | -1.3038619 |
| 1.0253472 | -1.0293786 |
| 0.90012265 | -0.90285366 |
| 0.75743853 | -0.75446879 |

## X. FIGURE 4: RAW DATA T=293 K, T=321 K

| T=293 K | | T=293 K | | T=321 K | | T=321 K | |
|---|---|---|---|---|---|---|---|
| Gd (MΩ)^(-1) | Voltage (V) | Gd (MΩ)^(-1) | Voltage (V) | Gd (MΩ)^(-1) | Voltage (V) | Gd (MΩ)^(-1) | Voltage (V) |
| 0 | 0.62366017 | 0.69455197 | 1.1325762 | 0 | 1.1268119 | 0.56338976 | 1.9640849 |
| 0.014175104 | 0.62366017 | 0.7023155 | 1.1387154 | 0.011247723 | 1.1268119 | 0.56981127 | 1.9736889 |
| 0.028350208 | 0.62366017 | 0.71003805 | 1.1447572 | 0.022495445 | 1.1268119 | 0.57620256 | 1.9830236 |
| 0.042525312 | 0.62366017 | 0.71772206 | 1.1504987 | 0.033743168 | 1.1268119 | 0.58256431 | 1.9922269 |
| 0.056700416 | 0.62366017 | 0.72536732 | 1.1563315 | 0.044990891 | 1.1268119 | 0.58889819 | 2.0009978 |
| 0.07087552 | 0.62366017 | 0.73297647 | 1.1618178 | 0.056238613 | 1.1268119 | 0.59520418 | 2.0098461 |
| 0.085050624 | 0.62366017 | 0.74054974 | 1.1673223 | 0.067486336 | 1.1268119 | 0.60148322 | 2.0184718 |
| 0.099225728 | 0.62366017 | 0.74808779 | 1.1727758 | 0.078734059 | 1.1268119 | 0.60773672 | 2.026716 |
| 0.11340083 | 0.62366017 | 0.75559285 | 1.1779327 | 0.089981781 | 1.1268119 | 0.61396514 | 2.0348747 |
| 0.12757594 | 0.62366017 | 0.76306512 | 1.1831002 | 0.1012295 | 1.1268119 | 0.62016873 | 2.0430222 |
| 0.14175104 | 0.62366017 | 0.77050483 | 1.1882778 | 0.11247723 | 1.1268119 | 0.62634889 | 2.0507685 |
| 0.15592614 | 0.62366017 | 0.77791398 | 1.1931798 | 0.12372495 | 1.1268119 | 0.63250608 | 2.0584172 |
| 0.17010125 | 0.62366017 | 0.78529301 | 1.1980505 | 0.13497267 | 1.1268119 | 0.63864101 | 2.0658863 |
| 0.18427635 | 0.62366017 | 0.79264302 | 1.2027799 | 0.14622039 | 1.1268119 | 0.64475391 | 2.0733304 |
| 0.19845146 | 0.62366017 | 0.79996358 | 1.2076198 | 0.15746812 | 1.1268119 | 0.65084479 | 2.0808283 |
| 0.21262545 | 0.62370889 | 0.80725711 | 1.2120939 | 0.16871584 | 1.1268119 | 0.65691598 | 2.0875746 |
| 0.22679856 | 0.62374788 | 0.81452296 | 1.2167124 | 0.17996356 | 1.1268119 | 0.66296678 | 2.0946081 |
| 0.24078961 | 0.63186453 | 0.8217629 | 1.2210674 | 0.19121129 | 1.1268119 | 0.66899861 | 2.1011992 |
| 0.25471687 | 0.63475852 | 0.82897692 | 1.2254536 | 0.20245831 | 1.1268823 | 0.67501052 | 2.1081606 |
| 0.26846097 | 0.64321801 | 0.83616635 | 1.2296442 | 0.21370439 | 1.1269762 | 0.68100485 | 2.1143415 |
| 0.28196962 | 0.65442853 | 0.84333098 | 1.2339017 | 0.22466178 | 1.1566684 | 0.6869802 | 2.1210576 |
| 0.29520208 | 0.66808802 | 0.85047103 | 1.2381503 | 0.23558918 | 1.1598432 | 0.69293775 | 2.1273981 |
| 0.30813354 | 0.68363879 | 0.85758848 | 1.2420803 | 0.24632372 | 1.1806804 | 0.69887819 | 2.1335241 |
| 0.32075403 | 0.70048351 | 0.86468356 | 1.2459965 | 0.25682792 | 1.2065712 | 0.70480199 | 2.1395162 |
| 0.33306511 | 0.71808885 | 0.87175472 | 1.2502114 | 0.26707741 | 1.2365562 | 0.71070938 | 2.1454569 |
| 0.34507785 | 0.73592291 | 0.87880617 | 1.2537064 | 0.27706328 | 1.2692003 | 0.71660132 | 2.1510885 |
| 0.35680863 | 0.75361096 | 0.88583414 | 1.2578945 | 0.2867874 | 1.3033633 | 0.72247638 | 2.1572658 |
| 0.36827651 | 0.77088767 | 0.89284351 | 1.2612333 | 0.29626127 | 1.337793 | 0.72833714 | 2.1625273 |
| 0.3795012 | 0.78758955 | 0.89983095 | 1.2651912 | 0.30550104 | 1.3716865 | 0.73418127 | 2.1686836 |
| 0.39050197 | 0.80362106 | 0.90679846 | 1.2688108 | 0.3145264 | 1.4042728 | 0.74001158 | 2.1738262 |
| 0.40129675 | 0.81895545 | 0.91374913 | 1.2718836 | 0.32335633 | 1.4353529 | 0.74582805 | 2.1789932 |
| 0.4119015 | 0.83363097 | 0.92067876 | 1.2757455 | 0.33200981 | 1.4646205 | 0.75163094 | 2.1840966 |
| 0.42233061 | 0.84767028 | 0.92758979 | 1.2791799 | 0.34050395 | 1.4920955 | 0.7574193 | 2.1895785 |
| 0.43259627 | 0.86116746 | 0.93448177 | 1.2827153 | 0.34885421 | 1.5178047 | 0.76319407 | 2.1947318 |
| 0.4427091 | 0.87418145 | 0.9413567 | 1.2858973 | 0.35707348 | 1.5419937 | 0.76895525 | 2.1999093 |
| 0.45267862 | 0.88674688 | 0.94821479 | 1.2890534 | 0.36517325 | 1.5647448 | 0.77470518 | 2.2042126 |
| 0.46251305 | 0.89892912 | 0.95505561 | 1.2923088 | 0.37316335 | 1.5862212 | 0.78044222 | 2.2091643 |
| 0.47221989 | 0.91074364 | 0.96188003 | 1.2954125 | 0.38105175 | 1.6066709 | 0.78616543 | 2.2145009 |
| 0.48180603 | 0.92221184 | 0.96868674 | 1.2987847 | 0.38884572 | 1.6261378 | 0.79187717 | 2.2189526 |
| 0.49127699 | 0.9334265 | 0.97547927 | 1.3014951 | 0.39655088 | 1.6448808 | 0.79757766 | 2.2233308 |
| 0.50063876 | 0.94431368 | 0.9822532 | 1.3050697 | 0.40417285 | 1.662833 | 0.80326619 | 2.2280017 |
| 0.50989643 | 0.95493207 | 0.98901384 | 1.307635 | 0.41171586 | 1.6802414 | 0.80894348 | 2.2324158 |
| 0.51905488 | 0.96527813 | 0.99575964 | 1.3105116 | 0.41918411 | 1.6970591 | 0.81460952 | 2.2368474 |
| 0.52811831 | 0.97539773 | 1.0024884 | 1.3138331 | 0.42658113 | 1.7134023 | 0.82026408 | 2.2413895 |
| 0.53709159 | 0.98519649 | 1.0092036 | 1.3164765 | 0.43391019 | 1.7292889 | 0.82590739 | 2.2458568 |
| 0.54597805 | 0.99482206 | 1.0159029 | 1.3196102 | 0.44117458 | 1.7446847 | 0.83154226 | 2.249219 |
| 0.5547819 | 1.0041574 | 1.0225905 | 1.3219266 | 0.44837687 | 1.759727 | 0.83716542 | 2.2539055 |
| 0.56350645 | 1.0132835 | 1.0292647 | 1.3245587 | 0.45551988 | 1.7743322 | 0.8427785 | 2.2579515 |
| 0.57215504 | 1.0221843 | 1.0359226 | 1.3278194 | 0.46260595 | 1.7885903 | 0.84838338 | 2.2612555 |
| 0.58073098 | 1.0308432 | 1.0425703 | 1.3298544 | 0.46963741 | 1.8024784 | 0.85397561 | 2.2663721 |
| 0.58923715 | 1.0392983 | 1.0492022 | 1.3330077 | 0.47661639 | 1.816035 | 0.85955893 | 2.2699865 |
| 0.59767621 | 1.0475631 | 1.0558227 | 1.3353267 | 0.48354522 | 1.8291782 | 0.86513265 | 2.2738993 |
| 0.60605104 | 1.0555974 | 1.0624287 | 1.3382368 | 0.49042555 | 1.8420737 | 0.8706991 | 2.2768667 |
| 0.61436385 | 1.0634725 | 1.0690239 | 1.340439 | 0.49725948 | 1.8545799 | 0.87625524 | 2.2810919 |
| 0.62261752 | 1.0710922 | 1.0756076 | 1.3427839 | 0.50404888 | 1.8667415 | 0.8818006 | 2.2855259 |
| 0.63081361 | 1.0786178 | 1.0821773 | 1.3456359 | 0.51079494 | 1.8787373 | 0.88733964 | 2.2881364 |
| 0.63895433 | 1.0859543 | 1.0887361 | 1.3478624 | 0.51749975 | 1.8902935 | 0.89287 | 2.2917236 |
| 0.64704211 | 1.093062 | 1.0952824 | 1.3504619 | 0.5241645 | 1.9016584 | | |
| 0.65507873 | 1.1000207 | 1.10182 | 1.3522462 | 0.53079081 | 1.9126872 | | |
| 0.66306574 | 1.1068537 | 1.1083454 | 1.3547706 | 0.53737987 | 1.9235026 | | |
| 0.67100491 | 1.1135235 | 1.1148578 | 1.3574891 | 0.54393331 | 1.9339568 | | |
| 0.678898 | 1.1200227 | 1.1213613 | 1.3593383 | 0.55045183 | 1.9443155 | | |
| 0.68674636 | 1.1264075 | 1.1278539 | 1.3616105 | 0.55693685 | 1.954362 | | |

## XI.  FIGURE 4: RAW DATA T=344 K, T=370 K

| T=344 K | | T=344 K | | T=370 K | | T=370 K | |
|---|---|---|---|---|---|---|---|
| Gd (MΩ)^(-1) | Voltage (V) | Gd (MΩ)^(-1) | Voltage (V) | Gd (MΩ)^(-1) | Voltage (V) | Gd (MΩ)^(-1) | Voltage (V) |
| 0 | 1.6028457 | 0.64848163 | 2.9248533 | 0 | 2.4055267 | 0.40836603 | 3.630527 |
| 0.01474063 | 1.6028457 | 0.65654226 | 2.9311572 | 0.0079643337 | 2.4055267 | 0.41363006 | 3.6394987 |
| 0.02948126 | 1.6028457 | 0.66458511 | 2.9376325 | 0.015928667 | 2.4055267 | 0.41888134 | 3.6483304 |
| 0.044221889 | 1.6028457 | 0.67261204 | 2.9434606 | 0.023893001 | 2.4055267 | 0.42412055 | 3.6567418 |
| 0.058962519 | 1.6028457 | 0.68062305 | 2.949312 | 0.031857335 | 2.4055267 | 0.4293478 | 3.665099 |
| 0.073703149 | 1.6028457 | 0.68861629 | 2.9558678 | 0.039821668 | 2.4055267 | 0.43456418 | 3.6727467 |
| 0.088443779 | 1.6028457 | 0.69659334 | 2.9618664 | 0.047786002 | 2.4055267 | 0.439769 | 3.6808957 |
| 0.10318441 | 1.6028457 | 0.70455696 | 2.9668593 | 0.055750336 | 2.4055267 | 0.44496307 | 3.6885152 |
| 0.11792464 | 1.6028887 | 0.71250572 | 2.9724103 | 0.063714669 | 2.4055267 | 0.45014719 | 3.6955985 |
| 0.13256459 | 1.613869 | 0.72044118 | 2.9773895 | 0.071679003 | 2.4055267 | 0.45532056 | 3.7032791 |
| 0.14696697 | 1.6404891 | 0.72835966 | 2.9837733 | 0.079643337 | 2.4055267 | 0.4604845 | 3.7100378 |
| 0.16094267 | 1.6905745 | 0.7362688 | 2.9872986 | 0.08760767 | 2.4055267 | 0.46563888 | 3.7169169 |
| 0.17439323 | 1.7565776 | 0.74416319 | 2.9928765 | 0.095572004 | 2.4055267 | 0.47078384 | 3.7237255 |
| 0.18731195 | 1.8288937 | 0.75204456 | 2.9978244 | 0.10353634 | 2.4055267 | 0.47591978 | 3.7302698 |
| 0.19974725 | 1.8999907 | 0.75991487 | 3.0020355 | 0.11146563 | 2.4161578 | 0.48104668 | 3.7368372 |
| 0.21176468 | 1.9660572 | 0.7677736 | 3.0064598 | 0.11939757 | 2.4153491 | 0.48616563 | 3.7426512 |
| 0.22342754 | 2.0258277 | 0.77562049 | 3.0109982 | 0.12723713 | 2.4438131 | 0.49127607 | 3.7488727 |
| 0.23478888 | 2.0795922 | 0.78345316 | 3.0164623 | 0.13496439 | 2.479328 | 0.49637789 | 3.7552127 |
| 0.24589042 | 2.1282599 | 0.79127938 | 3.018948 | 0.14254923 | 2.5258851 | 0.50147215 | 3.76079 |
| 0.25676479 | 2.1727188 | 0.79909297 | 3.0238297 | 0.14997491 | 2.5800222 | 0.50655897 | 3.7662857 |
| 0.26743688 | 2.2139026 | 0.80689589 | 3.027961 | 0.15723758 | 2.6379281 | 0.51163795 | 3.7720932 |
| 0.27792694 | 2.2523176 | 0.81468908 | 3.0317451 | 0.1643427 | 2.6964261 | 0.51670991 | 3.7773253 |
| 0.28825144 | 2.2884369 | 0.82247068 | 3.0362574 | 0.171301 | 2.7533164 | 0.52177456 | 3.7827703 |
| 0.29842458 | 2.322484 | 0.83024215 | 3.0402168 | 0.17812577 | 2.8071886 | 0.52683204 | 3.7881316 |
| 0.30845873 | 2.3546527 | 0.83800704 | 3.0427933 | 0.18483055 | 2.8574291 | 0.53188342 | 3.7927106 |
| 0.31836457 | 2.3851554 | 0.84576259 | 3.0464595 | 0.19142807 | 2.9038811 | 0.53692763 | 3.7981001 |
| 0.32815208 | 2.4139892 | 0.85350471 | 3.051742 | 0.19792962 | 2.9467456 | 0.54196534 | 3.8030038 |
| 0.33782983 | 2.4413687 | 0.86124196 | 3.0536627 | 0.20434529 | 2.9861916 | 0.54699721 | 3.807418 |
| 0.34740585 | 2.4673061 | 0.8689725 | 3.0563141 | 0.21068344 | 3.0227146 | 0.55202297 | 3.8120437 |
| 0.35688723 | 2.4919312 | 0.87669198 | 3.0606912 | 0.21695137 | 3.0565777 | 0.55704236 | 3.8168826 |
| 0.36628017 | 2.5153951 | 0.88440383 | 3.0637208 | 0.22315519 | 3.0881657 | 0.56205551 | 3.8216327 |
| 0.37559072 | 2.537654 | 0.89210371 | 3.0684863 | 0.2293002 | 3.1177172 | 0.56706361 | 3.8254817 |
| 0.38482415 | 2.5588506 | 0.89980292 | 3.0687485 | 0.23539066 | 3.1456444 | 0.57206641 | 3.8295418 |
| 0.39398492 | 2.5791437 | 0.90748845 | 3.0742139 | 0.24143055 | 3.1719837 | 0.5770635 | 3.833916 |
| 0.40307805 | 2.5983315 | 0.91517056 | 3.0755833 | 0.24742318 | 3.1969964 | 0.58205395 | 3.8390148 |
| 0.41210682 | 2.6168528 | 0.92284555 | 3.0784313 | 0.25337134 | 3.2208967 | 0.58704055 | 3.8419783 |
| 0.42107544 | 2.6344025 | 0.93051252 | 3.0816548 | 0.25927756 | 3.2437712 | 0.59202224 | 3.8457661 |
| 0.42998707 | 2.651249 | 0.9381716 | 3.0848321 | 0.26514435 | 3.2655686 | 0.59699729 | 3.8508965 |
| 0.43884487 | 2.6673609 | 0.94582356 | 3.0876973 | 0.27097344 | 3.2866876 | 0.6019689 | 3.8535697 |
| 0.44765148 | 2.6828676 | 0.9534679 | 3.0907806 | 0.2767671 | 3.3067931 | 0.60693519 | 3.8576896 |
| 0.45640991 | 2.697623 | 0.96110578 | 3.0933903 | 0.28252677 | 3.3263027 | 0.61189684 | 3.8613018 |
| 0.46512189 | 2.712009 | 0.96874274 | 3.0937635 | 0.28825432 | 3.3449582 | 0.61685424 | 3.8646103 |
| 0.47379043 | 2.725597 | 0.97636957 | 3.0978744 | 0.29395121 | 3.3629619 | 0.62180646 | 3.8686502 |
| 0.48241738 | 2.7387368 | 0.98398521 | 3.102425 | 0.29961863 | 3.3804478 | 0.6267547 | 3.8717635 |
| 0.49100433 | 2.7514978 | 0.99160335 | 3.1014066 | 0.30525804 | 3.3972367 | 0.63169829 | 3.8754021 |
| 0.49955389 | 2.7635271 | 0.99920768 | 3.1070428 | 0.31087064 | 3.4134664 | 0.63663764 | 3.8787348 |
| 0.508067 | 2.7753617 | 1.0068107 | 3.1075807 | 0.31645775 | 3.4290371 | 0.6415746 | 3.880612 |
| 0.51654563 | 2.7866491 | 1.0144134 | 3.1076883 | 0.32202018 | 3.4442573 | 0.64650465 | 3.8860451 |
| 0.52499135 | 2.7975054 | 1.0219981 | 3.1150761 | 0.32755911 | 3.458867 | 0.65143258 | 3.8877199 |
| 0.53340456 | 2.8083149 | 1.0295917 | 3.1114587 | 0.33307547 | 3.473016 | 0.65635574 | 3.8914934 |
| 0.5417879 | 2.818323 | 1.0371662 | 3.1192438 | 0.3385702 | 3.4866916 | 0.66127491 | 3.8946437 |
| 0.55014242 | 2.8280463 | 1.044744 | 3.1179436 | 0.34404422 | 3.4998812 | 0.66619182 | 3.8964311 |
| 0.55846837 | 2.8377471 | 1.0523088 | 3.1232597 | 0.34949819 | 3.5127434 | 0.67110369 | 3.9004324 |
| 0.56676721 | 2.847018 | 1.0598838 | 3.1190812 | 0.35493279 | 3.5252699 | 0.67601251 | 3.9028582 |
| 0.57504052 | 2.8558044 | 1.0674373 | 3.1279398 | 0.36034906 | 3.5371924 | 0.68091681 | 3.9064498 |
| 0.58328975 | 2.8641425 | 1.0749982 | 3.1248907 | 0.36574756 | 3.5488468 | 0.68581873 | 3.9083539 |
| 0.59151436 | 2.8727132 | 1.082543 | 3.131541 | 0.37112879 | 3.5602269 | 0.69071653 | 3.9116375 |
| 0.59971581 | 2.8808268 | 1.0900913 | 3.1301214 | 0.37649357 | 3.57115 | 0.69561101 | 3.9142896 |
| 0.60789462 | 2.8888004 | 1.0976298 | 3.1341653 | 0.38184241 | 3.5817848 | 0.70050177 | 3.9172642 |
| 0.61605212 | 2.8963508 | 1.1051689 | 3.1339465 | 0.38717586 | 3.5921254 | 0.70538895 | 3.9201369 |
| 0.6241896 | 2.9034712 | 1.112703 | 3.1359721 | 0.39249431 | 3.6022562 | 0.71027401 | 3.9218412 |
| 0.63230616 | 2.910957 | 1.1202311 | 3.138494 | 0.39779856 | 3.6119019 | 0.71515628 | 3.9240804 |
| 0.64040364 | 2.9178175 | | | 0.403089 | 3.6213268 | 0.7200351 | 3.9268562 |

## XII. FIGURE 4: RAW DATA T=415 K, T=433 K

| T=415 K | | T=415 K | | T=433 K | | T=433 K | |
|---|---|---|---|---|---|---|---|
| Gd (MΩ)^(-1) | Voltage (V) | Gd (MΩ)^(-1) | Voltage (V) | Gd (MΩ)^(-1) | Voltage (V) | Gd (MΩ)^(-1) | Voltage (V) |
| 0 | 3.6320644 | 0.46178046 | 5.4998952 | 0 | 4.4645375 | 0.2484339 | 5.996261 |
| 0.0095210045 | 3.6320644 | 0.46805964 | 5.5072281 | 0.0045182435 | 4.4645375 | 0.25179112 | 6.00851 |
| 0.019042009 | 3.6320644 | 0.47433095 | 5.5141475 | 0.009036487 | 4.4645375 | 0.25514178 | 6.0202655 |
| 0.028563014 | 3.6320644 | 0.48059544 | 5.520144 | 0.013554731 | 4.4645375 | 0.25848599 | 6.0318852 |
| 0.038084018 | 3.6320644 | 0.48685173 | 5.527386 | 0.018072974 | 4.4645375 | 0.26182414 | 6.0428195 |
| 0.047605023 | 3.6320644 | 0.49310087 | 5.533702 | 0.022591217 | 4.4645375 | 0.26515615 | 6.0539767 |
| 0.057126027 | 3.6320644 | 0.49934287 | 5.5400325 | 0.027109461 | 4.4645375 | 0.2684824 | 6.0644396 |
| 0.066647032 | 3.6320644 | 0.50557938 | 5.5449176 | 0.031627704 | 4.4645375 | 0.27180301 | 6.0747542 |
| 0.076166887 | 3.6325028 | 0.51180841 | 5.5515663 | 0.036145948 | 4.4645375 | 0.27511796 | 6.085104 |
| 0.085566171 | 3.6790998 | 0.51803137 | 5.5569848 | 0.040664191 | 4.4645375 | 0.27842778 | 6.0945604 |
| 0.094801956 | 3.7442299 | 0.524249 | 5.5617529 | 0.045182435 | 4.4645375 | 0.28173225 | 6.1044188 |
| 0.10379143 | 3.8468233 | 0.53045949 | 5.5681478 | 0.049700678 | 4.4645375 | 0.28503178 | 6.1135616 |
| 0.11252046 | 3.9615939 | 0.53666521 | 5.5724192 | 0.054218922 | 4.4645375 | 0.28832646 | 6.1225445 |
| 0.12100729 | 4.0746564 | 0.54286544 | 5.5773616 | 0.058737165 | 4.4645375 | 0.29161631 | 6.1315537 |
| 0.12928195 | 4.1791348 | 0.54905901 | 5.5833484 | 0.063255409 | 4.4645375 | 0.29490152 | 6.1402125 |
| 0.13737455 | 4.273147 | 0.55524865 | 5.5869022 | 0.067773652 | 4.4645375 | 0.29818219 | 6.1487068 |
| 0.14531121 | 4.3571101 | 0.56143304 | 5.5916477 | 0.072291896 | 4.4645375 | 0.30145852 | 6.1568455 |
| 0.15311343 | 4.4321892 | 0.56761274 | 5.5958809 | 0.076810139 | 4.4645375 | 0.30473061 | 6.1648157 |
| 0.16079828 | 4.4998822 | 0.57378802 | 5.5998973 | 0.08132798 | 4.4649362 | 0.30799847 | 6.1728067 |
| 0.16837946 | 4.561413 | 0.57995689 | 5.6057088 | 0.085794082 | 4.5166606 | 0.3112623 | 6.1804364 |
| 0.17586789 | 4.6179082 | 0.5861224 | 5.6087685 | 0.090242434 | 4.5346834 | 0.31452219 | 6.1878935 |
| 0.18327269 | 4.6700674 | 0.5922824 | 5.6137756 | 0.094642578 | 4.5843655 | 0.31777835 | 6.1949849 |
| 0.19060124 | 4.7186573 | 0.59843921 | 5.6166943 | 0.098984529 | 4.645807 | 0.32103068 | 6.2022849 |
| 0.19786002 | 4.7640905 | 0.60459282 | 5.619616 | 0.10326264 | 4.7151343 | 0.32427938 | 6.2092165 |
| 0.20505461 | 4.8065103 | 0.61074035 | 5.6251682 | 0.10747499 | 4.7887395 | 0.32752434 | 6.2163569 |
| 0.21218995 | 4.8464291 | 0.61688591 | 5.6269712 | 0.1116228 | 4.8632597 | 0.33076608 | 6.2225456 |
| 0.21927054 | 4.8839008 | 0.62302565 | 5.632312 | 0.11570929 | 4.9362344 | 0.33400429 | 6.2293286 |
| 0.22630016 | 4.9193129 | 0.62916234 | 5.6350993 | 0.1197388 | 5.006039 | 0.33723907 | 6.235932 |
| 0.23328234 | 4.9527375 | 0.63529543 | 5.6384175 | 0.12371596 | 5.071923 | 0.34047073 | 6.2419649 |
| 0.24022053 | 4.9841418 | 0.6414263 | 5.6404556 | 0.12764572 | 5.133101 | 0.34369925 | 6.2480096 |
| 0.24711751 | 5.0139173 | 0.64755192 | 5.6452925 | 0.13153242 | 5.1899757 | 0.34692465 | 6.2540659 |
| 0.253976 | 5.0420587 | 0.6536731 | 5.6493801 | 0.13538009 | 5.2426223 | 0.35014713 | 6.2597422 |
| 0.26079829 | 5.0688098 | 0.6597938 | 5.6498347 | 0.13919236 | 5.2913038 | 0.35336638 | 6.2660176 |
| 0.26758685 | 5.0939979 | 0.66590825 | 5.6555986 | 0.14297255 | 5.3361956 | 0.35658311 | 6.2709291 |
| 0.27434356 | 5.1180072 | 0.67202328 | 5.6550672 | 0.1467235 | 5.3778039 | 0.35979671 | 6.2770299 |
| 0.28107031 | 5.1408008 | 0.67813232 | 5.6606136 | 0.15044752 | 5.4166901 | 0.36300758 | 6.2823533 |
| 0.28776859 | 5.1626594 | 0.68423922 | 5.6625917 | 0.15414674 | 5.453019 | 0.36621604 | 6.2870928 |
| 0.29444043 | 5.1831101 | 0.6903426 | 5.6658661 | 0.15782325 | 5.4866759 | 0.36942177 | 6.2924332 |
| 0.30108708 | 5.2027596 | 0.69644474 | 5.6670093 | 0.16147849 | 5.5186183 | 0.37262468 | 6.297981 |
| 0.30770984 | 5.221523 | 0.70254171 | 5.6718155 | 0.16511386 | 5.5487788 | 0.37582538 | 6.3023469 |
| 0.31430986 | 5.2395099 | 0.70863655 | 5.6738014 | 0.16873078 | 5.5770929 | 0.37902345 | 6.3075144 |
| 0.32088863 | 5.2564404 | 0.71472901 | 5.6760181 | 0.17233024 | 5.6041267 | 0.38221879 | 6.3128896 |
| 0.32744688 | 5.2728868 | 0.7208204 | 5.6770123 | 0.17591327 | 5.6298388 | 0.38541222 | 6.3166777 |
| 0.33398584 | 5.2884404 | 0.72690801 | 5.6805332 | 0.17948077 | 5.6543489 | 0.38860303 | 6.3218687 |
| 0.34050353 | 5.3034184 | 0.73299308 | 5.6829085 | 0.18303354 | 5.6777836 | 0.39179192 | 6.3256675 |
| 0.34700918 | 5.3178102 | 0.73907668 | 5.6842886 | 0.18657229 | 5.7002759 | 0.39497829 | 6.330673 |
| 0.3534952 | 5.3316056 | 0.74515715 | 5.6872043 | 0.19009773 | 5.7218012 | 0.39816254 | 6.3348837 |
| 0.35996554 | 5.3445234 | 0.75123328 | 5.691276 | 0.19361026 | 5.74283 | 0.40134468 | 6.3391 |
| 0.36642021 | 5.357504 | 0.75730842 | 5.6921987 | 0.19711069 | 5.7626853 | 0.404525 | 6.3427184 |
| 0.37286018 | 5.3697264 | 0.76338397 | 5.6918142 | 0.20059942 | 5.7820098 | 0.40770279 | 6.3477509 |
| 0.3792867 | 5.3809736 | 0.76945443 | 5.6965856 | 0.20407696 | 5.800623 | 0.41087908 | 6.3507742 |
| 0.38569934 | 5.3926131 | 0.77552218 | 5.6991285 | 0.2075436 | 5.8188486 | 0.41405254 | 6.3564254 |
| 0.39209926 | 5.4033328 | 0.7815887 | 5.7002851 | 0.21099995 | 5.8361671 | 0.41722509 | 6.358244 |
| 0.39848704 | 5.4136082 | 0.78765407 | 5.701365 | 0.21444643 | 5.8529038 | 0.42039513 | 6.3633012 |
| 0.40486274 | 5.4238529 | 0.79371657 | 5.7040666 | 0.21788312 | 5.8695646 | 0.42356344 | 6.3667446 |
| 0.41122671 | 5.4338562 | 0.79977325 | 5.7095548 | 0.22131073 | 5.8851081 | 0.42672994 | 6.3703947 |
| 0.41758051 | 5.4425602 | 0.80583583 | 5.7039894 | 0.22472926 | 5.9007341 | 0.42989473 | 6.3738458 |
| 0.42392289 | 5.4523504 | 0.81188889 | 5.7129613 | 0.22813923 | 5.9155684 | 0.4330572 | 6.378521 |
| 0.43025633 | 5.4600522 | 0.81794236 | 5.712574 | 0.23154092 | 5.92995 | 0.43621866 | 6.3805558 |
| 0.43657902 | 5.4693374 | 0.82399813 | 5.7104061 | 0.23493454 | 5.9440483 | 0.4393782 | 6.3844255 |
| 0.44289267 | 5.4771586 | 0.83004848 | 5.7155188 | 0.2383204 | 5.9576815 | 0.44253573 | 6.388504 |
| 0.44919697 | 5.4852878 | 0.83609522 | 5.7189324 | 0.2416988 | 5.9708425 | 0.44569204 | 6.3909535 |
| 0.4554929 | 5.4925818 | 0.8421482 | 5.7130388 | 0.24506983 | 5.9838828 | 0.44884695 | 6.3938138 |

# XIII. FIGURE 4: RAW DATA T=454 K, T=470 K

| T=454 K | | T=454 K | | T=470 K | | T=470 K | |
|---|---|---|---|---|---|---|---|
| Gd (MΩ)^(-1) | Voltage (V) | Gd (MΩ)^(-1) | Voltage (V) | Gd (MΩ)^(-1) | Voltage (V) | Gd (MΩ)^(-1) | Voltage (V) |
| 0 | 5.8509396 | 0.26115152 | 7.5586722 | 0 | 7.4900838 | 0.26863804 | 8.9196866 |
| 0.0047695784 | 5.8509396 | 0.26483944 | 7.5670079 | 0.0046 | 7.4900838 | 0.27249549 | 8.9319188 |
| 0.0095391568 | 5.8509396 | 0.26852357 | 7.574792 | 0.0092 | 7.4900838 | 0.27634788 | 8.9436507 |
| 0.014308735 | 5.8509396 | 0.27220391 | 7.582592 | 0.0138 | 7.4900838 | 0.28019543 | 8.954878 |
| 0.019078314 | 5.8509396 | 0.27588055 | 7.5902173 | 0.0184 | 7.4900838 | 0.28403804 | 8.9664019 |
| 0.023847892 | 5.8509396 | 0.27955386 | 7.5970932 | 0.023 | 7.4900838 | 0.28787617 | 8.9768795 |
| 0.02861747 | 5.8509396 | 0.28322357 | 7.6045561 | 0.0276 | 7.4900838 | 0.29170969 | 8.9876512 |
| 0.033387049 | 5.8509396 | 0.28689014 | 7.6110743 | 0.0322 | 7.4900838 | 0.29553896 | 8.9976381 |
| 0.038156627 | 5.8509396 | 0.29055347 | 7.6177958 | 0.0368 | 7.4900838 | 0.29936386 | 9.007918 |
| 0.042926205 | 5.8509396 | 0.29421366 | 7.6243367 | 0.0414 | 7.4900838 | 0.30318474 | 9.0174072 |
| 0.047695784 | 5.8509396 | 0.29787089 | 7.630503 | 0.046 | 7.4900838 | 0.30700159 | 9.0269163 |
| 0.052465362 | 5.8509396 | 0.30152507 | 7.6368726 | 0.0506 | 7.4900838 | 0.31081453 | 9.036173 |
| 0.057234941 | 5.8509396 | 0.30517648 | 7.6426723 | 0.0552 | 7.4900838 | 0.31462367 | 9.0451757 |
| 0.062004519 | 5.8509396 | 0.30882521 | 7.6482871 | 0.0598 | 7.4900838 | 0.31842914 | 9.0539226 |
| 0.066774097 | 5.8509396 | 0.31247116 | 7.6541041 | 0.0644 | 7.4900838 | 0.32223103 | 9.0624124 |
| 0.071538869 | 5.8568418 | 0.31611462 | 7.6593471 | 0.069 | 7.4900838 | 0.32602949 | 9.0706435 |
| 0.076238568 | 5.9379373 | 0.31975549 | 7.6647918 | 0.0736 | 7.4900838 | 0.32982449 | 9.0788895 |
| 0.080897503 | 5.9898911 | 0.32339396 | 7.6698545 | 0.0782 | 7.4900838 | 0.33361615 | 9.0868749 |
| 0.085483046 | 6.0857603 | 0.32703012 | 7.6747288 | 0.0828 | 7.4900838 | 0.3374046 | 9.0945983 |
| 0.089988634 | 6.1937568 | 0.33066387 | 7.6798047 | 0.0874 | 7.4900838 | 0.34118993 | 9.1020584 |
| 0.094415468 | 6.3039439 | 0.33429531 | 7.6846917 | 0.09199655 | 7.4957056 | 0.34497229 | 9.1092537 |
| 0.098769372 | 6.4095383 | 0.33792481 | 7.6888016 | 0.096598735 | 7.4865277 | 0.34875153 | 9.1167378 |
| 0.10305811 | 6.5069288 | 0.34155191 | 7.6938961 | 0.10119529 | 7.4957056 | 0.35252779 | 9.1239563 |
| 0.10728991 | 6.59448 | 0.34517698 | 7.6982121 | 0.10566212 | 7.5000011 | 0.3563014 | 9.1303515 |
| 0.11147226 | 6.6724531 | 0.34880001 | 7.702533 | 0.11016552 | 7.6507496 | 0.36007213 | 9.1373129 |
| 0.11561171 | 6.7415868 | 0.352421 | 7.7068587 | 0.11463396 | 7.7106072 | 0.36384022 | 9.1437268 |
| 0.11971392 | 6.8028049 | 0.35604015 | 7.7107954 | 0.11906571 | 7.7744337 | 0.36760555 | 9.1504292 |
| 0.12378341 | 6.8575043 | 0.35965755 | 7.714539 | 0.12346009 | 7.8405567 | 0.37136835 | 9.1565817 |
| 0.12782415 | 6.9062904 | 0.36327318 | 7.7182862 | 0.12781813 | 7.9059361 | 0.37512862 | 9.1627425 |
| 0.13183928 | 6.9503311 | 0.3668866 | 7.7230247 | 0.13214121 | 7.96987 | 0.37888648 | 9.1686309 |
| 0.1358314 | 6.9904023 | 0.37049872 | 7.7257916 | 0.13643129 | 8.0311849 | 0.38264191 | 9.174527 |
| 0.13980281 | 7.026847 | 0.374109 | 7.7297476 | 0.14069054 | 8.0892986 | 0.38639494 | 9.1804306 |
| 0.14375528 | 7.0605352 | 0.37771752 | 7.7335096 | 0.14492116 | 8.1440511 | 0.39014578 | 9.1857786 |
| 0.14769036 | 7.0917148 | 0.38132465 | 7.7364822 | 0.14912522 | 8.1955126 | 0.39389432 | 9.1914147 |
| 0.15160946 | 7.120651 | 0.38492984 | 7.7406477 | 0.15330489 | 8.2433169 | 0.39764056 | 9.1970578 |
| 0.15551367 | 7.1477931 | 0.38853374 | 7.7434272 | 0.1574618 | 8.2884708 | 0.40138485 | 9.2018598 |
| 0.15940412 | 7.1730971 | 0.39213616 | 7.7466062 | 0.16159777 | 8.3304144 | 0.40512683 | 9.2075157 |
| 0.16328162 | 7.1970365 | 0.39573682 | 7.7503846 | 0.16571442 | 8.3695101 | 0.40886686 | 9.2123287 |
| 0.16714701 | 7.2195821 | 0.39933628 | 7.752972 | 0.16981291 | 8.406615 | 0.41260505 | 9.2168631 |
| 0.17100094 | 7.2410535 | 0.40293444 | 7.7557604 | 0.17389461 | 8.4411955 | 0.41634106 | 9.2222536 |
| 0.17484396 | 7.2616048 | 0.40653122 | 7.7587501 | 0.17796066 | 8.4736644 | 0.42007511 | 9.227082 |
| 0.17867682 | 7.2808683 | 0.41012625 | 7.7625404 | 0.182012 | 8.5044524 | 0.42380732 | 9.231631 |
| 0.18249988 | 7.2995281 | 0.41372025 | 7.7647365 | 0.18604965 | 8.5332769 | 0.42753792 | 9.2356151 |
| 0.18631378 | 7.3170423 | 0.41731306 | 7.7673334 | 0.1900743 | 8.5608296 | 0.43126645 | 9.2407425 |
| 0.19011881 | 7.3341062 | 0.42090466 | 7.7699321 | 0.19408665 | 8.5870838 | 0.43499337 | 9.2447344 |
| 0.19391544 | 7.350355 | 0.4244946 | 7.7735332 | 0.1980875 | 8.6117664 | 0.43871845 | 9.2493009 |
| 0.19770392 | 7.3661368 | 0.42808399 | 7.7747343 | 0.20207743 | 8.6353467 | 0.44244192 | 9.2533002 |
| 0.20148483 | 7.3809037 | 0.4316718 | 7.7781395 | 0.206057 | 8.6578053 | 0.44616343 | 9.2581612 |
| 0.20525834 | 7.3953675 | 0.43525841 | 7.7807454 | 0.21002669 | 8.6793752 | 0.44988357 | 9.2615955 |
| 0.20902483 | 7.4091608 | 0.43884392 | 7.7831524 | 0.21398694 | 8.7000422 | 0.45360221 | 9.2653189 |
| 0.21278457 | 7.4224581 | 0.4424284 | 7.7853602 | 0.21793823 | 8.7197926 | 0.45731924 | 9.2693321 |
| 0.21653775 | 7.435437 | 0.44601224 | 7.7867658 | 0.22188089 | 8.7388681 | 0.46103454 | 9.2736358 |
| 0.22028473 | 7.4477264 | 0.44959423 | 7.7907846 | 0.2258155 | 8.7567473 | 0.46474835 | 9.277369 |
| 0.2240258 | 7.4595035 | 0.45317567 | 7.791991 | 0.22974217 | 8.7744429 | 0.4684609 | 9.2805301 |
| 0.22776096 | 7.4713179 | 0.45675581 | 7.7948075 | 0.23366126 | 8.7914362 | 0.47217195 | 9.2842688 |
| 0.23149075 | 7.482057 | 0.46033521 | 7.7964178 | 0.23757321 | 8.8074596 | 0.4758815 | 9.2880105 |
| 0.23521509 | 7.4930131 | 0.4639136 | 7.7986331 | 0.24147815 | 8.8232817 | 0.47958979 | 9.2911789 |
| 0.23893425 | 7.5034418 | 0.46749078 | 7.8012528 | 0.24537642 | 8.8383785 | 0.48329647 | 9.2952145 |
| 0.2426486 | 7.5131516 | 0.47106695 | 7.8034708 | 0.24926825 | 8.8530038 | 0.487002 | 9.2980992 |
| 0.24635806 | 7.523074 | 0.47464265 | 7.8044795 | 0.25315376 | 8.8674151 | 0.49070615 | 9.3015633 |
| 0.25006299 | 7.5322709 | 0.47821679 | 7.8079107 | 0.25703339 | 8.8808203 | 0.49440915 | 9.304452 |
| 0.25376338 | 7.5414904 | 0.48179073 | 7.8083146 | 0.26090717 | 8.8942661 | 0.49811054 | 9.3084991 |
| 0.25745953 | 7.5501659 | 0.48536357 | 7.8107387 | 0.26477531 | 8.907223 | 0.50181101 | 9.3108134 |

# XIV. FIGURE 4: RAW DATA T=474 K

| T=474 K | | | T=474 K | |
|---|---|---|---|---|
| Gd (MΩ)^(-1) | Voltage (V) | | Gd (MΩ)^(-1) | Voltage (V) |
| 0 | 9.1700021 | | 0.2941383 | 9.8946602 |
| 0.0048 | 9.1700021 | | 0.29858097 | 9.9075556 |
| 0.0096 | 9.1700021 | | 0.30301794 | 9.9202834 |
| 0.0144 | 9.1700021 | | 0.30744939 | 9.9326406 |
| 0.0192 | 9.1700021 | | 0.31187544 | 9.9447589 |
| 0.024 | 9.1700021 | | 0.31629621 | 9.9566365 |
| 0.0288 | 9.1700021 | | 0.32071176 | 9.9684071 |
| 0.0336 | 9.1700021 | | 0.32512227 | 9.9797983 |
| 0.0384 | 9.1700021 | | 0.32952783 | 9.9910114 |
| 0.0432 | 9.1700021 | | 0.33392853 | 10.002045 |
| 0.048 | 9.1700021 | | 0.33832449 | 10.01283 |
| 0.0528 | 9.1700021 | | 0.34271577 | 10.023501 |
| 0.0576 | 9.1700021 | | 0.34710258 | 10.033715 |
| 0.0624 | 9.1700021 | | 0.35148492 | 10.043949 |
| 0.06719985 | 9.1702866 | | 0.35586294 | 10.05386 |
| 0.07199682 | 9.1757922 | | 0.36023664 | 10.06379 |
| 0.07679673 | 9.1701719 | | 0.36460626 | 10.073187 |
| 0.08159976 | 9.1642151 | | 0.36897177 | 10.082671 |
| 0.08639805 | 9.173268 | | 0.37333326 | 10.091964 |
| 0.09119472 | 9.1763661 | | 0.37769085 | 10.100996 |
| 0.095979 | 9.2001304 | | 0.38204469 | 10.109696 |
| 0.10073463 | 9.2055556 | | 0.38639472 | 10.118551 |
| 0.10550919 | 9.21886 | | 0.39074115 | 10.126932 |
| 0.11027769 | 9.2305757 | | 0.39508395 | 10.135397 |
| 0.11503956 | 9.2434275 | | 0.39942324 | 10.143595 |
| 0.11979432 | 9.2572496 | | 0.40375911 | 10.151596 |
| 0.12454149 | 9.2720505 | | 0.40809165 | 10.159398 |
| 0.12928068 | 9.2876631 | | 0.4124208 | 10.167354 |
| 0.13401156 | 9.3039773 | | 0.41674677 | 10.174828 |
| 0.13873383 | 9.320941 | | 0.42106956 | 10.182313 |
| 0.14344731 | 9.3383233 | | 0.4253892 | 10.189738 |
| 0.14815173 | 9.3563075 | | 0.42970584 | 10.19682 |
| 0.15284703 | 9.3744809 | | 0.43401945 | 10.203982 |
| 0.15753306 | 9.3930257 | | 0.43833027 | 10.210586 |
| 0.16220976 | 9.4117647 | | 0.44263809 | 10.217697 |
| 0.16687713 | 9.4305787 | | 0.44694315 | 10.224248 |
| 0.17153511 | 9.4495897 | | 0.45124545 | 10.230807 |
| 0.17618376 | 9.4685554 | | 0.45554499 | 10.237374 |
| 0.18082314 | 9.4874746 | | 0.45984192 | 10.243593 |
| 0.18545328 | 9.506408 | | 0.4641363 | 10.249675 |
| 0.19007427 | 9.5252316 | | 0.4684281 | 10.255837 |
| 0.19468623 | 9.5438816 | | 0.47271735 | 10.261934 |
| 0.19928922 | 9.56248 | | 0.4770042 | 10.267679 |
| 0.20388342 | 9.5807758 | | 0.48128865 | 10.273431 |
| 0.20846892 | 9.5989532 | | 0.48557055 | 10.279549 |
| 0.21304587 | 9.6168846 | | 0.48985029 | 10.284737 |
| 0.21761439 | 9.63463 | | 0.49412775 | 10.290219 |
| 0.22217466 | 9.6520601 | | 0.49840302 | 10.29549 |
| 0.22672683 | 9.6692347 | | 0.50267592 | 10.301201 |
| 0.23127102 | 9.6862147 | | 0.50694687 | 10.305904 |
| 0.23580738 | 9.7029336 | | 0.51121563 | 10.311191 |
| 0.24033612 | 9.7192597 | | 0.51548208 | 10.316774 |
| 0.24485733 | 9.7354469 | | 0.51974685 | 10.320838 |
| 0.24937119 | 9.7512993 | | 0.52400943 | 10.326141 |
| 0.25387788 | 9.7668133 | | 0.52826997 | 10.331085 |
| 0.25837749 | 9.7821811 | | 0.53252856 | 10.335815 |
| 0.26287023 | 9.7971394 | | 0.53678553 | 10.339749 |
| 0.26735622 | 9.811881 | | 0.54104028 | 10.345144 |
| 0.27183558 | 9.8264038 | | 0.54529335 | 10.34923 |
| 0.27630849 | 9.8405736 | | 0.54954468 | 10.353466 |
| 0.2807751 | 9.8544534 | | 0.55379418 | 10.357924 |
| 0.2852355 | 9.8681733 | | 0.5580417 | 10.362753 |
| 0.28968984 | 9.8815986 | | 0.56228775 | 10.36634 |

## XV. FIGURE 6: RAW DATA ICDW1 AND ICDW2

| ICDW1 | | ICDW1 | | ICDW2 | | ICDW2 | |
|---|---|---|---|---|---|---|---|
| I_CDW | Φ0/Φ | I_CDW | Φ0/Φ | I_CDW | Φ0/Φ | I_CDW | Φ0/Φ |
| -2.039 | 1.1320222 | 0.001 | 1.1328453 | -2.039 | 0.16281298 | 0.001 | 0.0045640661 |
| -1.999 | 1.1328453 | 0.041 | 1.1318339 | -1.999 | 0.0045640661 | 0.041 | 0.1704141 |
| -1.959 | 1.1318339 | 0.081 | 1.128778 | -1.959 | 0.1704141 | 0.081 | 0.30788361 |
| -1.919 | 1.128778 | 0.121 | 1.1243148 | -1.919 | 0.30788361 | 0.121 | 0.42304578 |
| -1.879 | 1.1243148 | 0.161 | 1.1180533 | -1.879 | 0.42304578 | 0.161 | 0.52089856 |
| -1.839 | 1.1180533 | 0.201 | 1.1096672 | -1.839 | 0.52089856 | 0.201 | 0.60459999 |
| -1.799 | 1.1096672 | 0.241 | 1.0989743 | -1.799 | 0.60459999 | 0.241 | 0.67716531 |
| -1.759 | 1.0989743 | 0.281 | 1.0863286 | -1.759 | 0.67716531 | 0.281 | 0.7399372 |
| -1.719 | 1.0863286 | 0.321 | 1.0713422 | -1.719 | 0.7399372 | 0.321 | 0.79506276 |
| -1.679 | 1.0713422 | 0.361 | 1.0540218 | -1.679 | 0.79506276 | 0.361 | 0.84355783 |
| -1.639 | 1.0540218 | 0.401 | 1.0339932 | -1.639 | 0.84355783 | 0.401 | 0.886487 |
| -1.599 | 1.0339932 | 0.441 | 1.0112948 | -1.599 | 0.886487 | 0.441 | 0.92397611 |
| -1.559 | 1.0112948 | 0.481 | 0.98513176 | -1.559 | 0.92397611 | 0.481 | 0.95706942 |
| -1.519 | 0.98513176 | 0.521 | 0.95566559 | -1.519 | 0.95706942 | 0.521 | 0.98628859 |
| -1.479 | 0.95566559 | 0.561 | 0.92216753 | -1.479 | 0.98628859 | 0.561 | 1.0123591 |
| -1.439 | 0.92216753 | 0.601 | 0.88418125 | -1.439 | 1.0123591 | 0.601 | 1.0352267 |
| -1.399 | 0.88418125 | 0.641 | 0.84129128 | -1.399 | 1.0352267 | 0.641 | 1.0547884 |
| -1.359 | 0.84129128 | 0.681 | 0.7925265 | -1.359 | 1.0547884 | 0.681 | 1.0723083 |
| -1.319 | 0.7925265 | 0.721 | 0.73709819 | -1.319 | 1.0723083 | 0.721 | 1.087021 |
| -1.279 | 0.73709819 | 0.761 | 0.67381722 | -1.279 | 1.087021 | 0.761 | 1.0995953 |
| -1.239 | 0.67381722 | 0.801 | 0.6008227 | -1.239 | 1.0995953 | 0.801 | 1.1099131 |
| -1.199 | 0.6008227 | 0.841 | 0.51637403 | -1.199 | 1.1099131 | 0.841 | 1.1184594 |
| -1.159 | 0.51637403 | 0.881 | 0.41774142 | -1.159 | 1.1184594 | 0.881 | 1.124956 |
| -1.119 | 0.41774142 | 0.921 | 0.30156795 | -1.119 | 1.124956 | 0.921 | 1.1292552 |
| -1.079 | 0.30156795 | 0.961 | 0.16281298 | -1.079 | 1.1292552 | 0.961 | 1.1320222 |
| -1.039 | 0.16281298 | 1.001 | 0.0045640661 | -1.039 | 1.1320222 | 1.001 | 1.1328453 |
| -0.999 | 0.0045640661 | 1.041 | 0.1704141 | -0.999 | 1.1328453 | 1.041 | 1.1318339 |
| -0.959 | 0.1704141 | 1.081 | 0.30788361 | -0.959 | 1.1318339 | 1.081 | 1.128778 |
| -0.919 | 0.30788361 | 1.121 | 0.42304578 | -0.919 | 1.128778 | 1.121 | 1.1243148 |
| -0.879 | 0.42304578 | 1.161 | 0.52089856 | -0.879 | 1.1243148 | 1.161 | 1.1180533 |
| -0.839 | 0.52089856 | 1.201 | 0.60459999 | -0.839 | 1.1180533 | 1.201 | 1.1096672 |
| -0.799 | 0.60459999 | 1.241 | 0.67716531 | -0.799 | 1.1096672 | 1.241 | 1.0989743 |
| -0.759 | 0.67716531 | 1.281 | 0.7399372 | -0.759 | 1.0989743 | 1.281 | 1.0863286 |
| -0.719 | 0.7399372 | 1.321 | 0.79506276 | -0.719 | 1.0863286 | 1.321 | 1.0713422 |
| -0.679 | 0.79506276 | 1.361 | 0.84355783 | -0.679 | 1.0713422 | 1.361 | 1.0540218 |
| -0.639 | 0.84355783 | 1.401 | 0.886487 | -0.639 | 1.0540218 | 1.401 | 1.0339932 |
| -0.599 | 0.886487 | 1.441 | 0.92397611 | -0.599 | 1.0339932 | 1.441 | 1.0112948 |
| -0.559 | 0.92397611 | 1.481 | 0.95706942 | -0.559 | 1.0112948 | 1.481 | 0.98513176 |
| -0.519 | 0.95706942 | 1.521 | 0.98628859 | -0.519 | 0.98513176 | 1.521 | 0.95566559 |
| -0.479 | 0.98628859 | 1.561 | 1.0123591 | -0.479 | 0.95566559 | 1.561 | 0.92216753 |
| -0.439 | 1.0123591 | 1.601 | 1.0352267 | -0.439 | 0.92216753 | 1.601 | 0.88418125 |
| -0.399 | 1.0352267 | 1.641 | 1.0547884 | -0.399 | 0.88418125 | 1.641 | 0.84129128 |
| -0.359 | 1.0547884 | 1.681 | 1.0723083 | -0.359 | 0.84129128 | 1.681 | 0.7925265 |
| -0.319 | 1.0723083 | 1.721 | 1.087021 | -0.319 | 0.7925265 | 1.721 | 0.73709819 |
| -0.279 | 1.087021 | 1.761 | 1.0995953 | -0.279 | 0.73709819 | 1.761 | 0.67381722 |
| -0.239 | 1.0995953 | 1.801 | 1.1099131 | -0.239 | 0.67381722 | 1.801 | 0.6008227 |
| -0.199 | 1.1099131 | 1.841 | 1.1184594 | -0.199 | 0.6008227 | 1.841 | 0.51637403 |
| -0.159 | 1.1184594 | 1.881 | 1.124956 | -0.159 | 0.51637403 | 1.881 | 0.41774142 |
| -0.119 | 1.124956 | 1.921 | 1.1292552 | -0.119 | 0.41774142 | 1.921 | 0.30156795 |
| -0.079 | 1.1292552 | 1.961 | 1.1320222 | -0.079 | 0.30156795 | 1.961 | 0.16281298 |
| -0.039 | 1.1320222 | 2.001 | 1.1328453 | -0.039 | 0.16281298 | 2.001 | 0.0045640661 |






% \end{thebibliography}
\end{document}